%% file: extspace.tex
\documentclass[12pt, utf8]{article}
\usepackage[body={17cm, 23cm, centered}]{geometry}
\usepackage[english]{babel}

\usepackage[utf8]{inputenc}

\usepackage{amsmath,amssymb,amsthm,amscd,cite,comment,amsfonts,indentfirst,color,setspace,bbold,arydshln}
\usepackage{mathtext,marvosym,textcomp}
\usepackage{lmodern}

\usepackage[makeroom]{cancel}
\usepackage{mathtools}
\usepackage{braket}
\usepackage{cancel}
\usepackage{float}
\usepackage{bbm}

\usepackage[debug,pageanchor=false]{hyperref}
\hypersetup{colorlinks=true,linktocpage,breaklinks,
	urlcolor=blue,
	linkcolor=blue,
	citecolor=blue
}

\usepackage[vcentermath]{youngtab}

\usepackage[mathmode,centertableaux]{ytableau}
\usepackage{tikz}
\usetikzlibrary{arrows}
\usetikzlibrary{shapes.misc}
\usepackage{tikz-cd}

\usetikzlibrary{arrows,shapes,positioning, fit}
\tikzstyle{every picture}+=[remember picture]
\tikzstyle{na} = [baseline=-.5ex]
\tikzstyle{format} = [rounded rectangle,
                      thick,
                      minimum size=1cm,
                      draw=red!00!black!80,
                      top color=white,
                      bottom color=red!00!black!00,
                      on grid]
\tikzstyle{format1} = [rectangle,
                      thick,
                      minimum size=1cm,
                      draw=red!00!black!80,
                      top color=white,
                      bottom color=red!00!black!00,
                      on grid]
\tikzstyle{format0} = [rounded rectangle,
					  thick,
				      minimum size=1.2cm,
					  draw=blue!00!black!80,
					  top color=white,
                      bottom color=blue!00!black!00,
                      text centered]
\tikzstyle{formatd} = [rounded rectangle,
					  thick,
					  dashed,
				      minimum size=1cm,
					  draw=blue!50!black!80,
					  top color=white,
                      bottom color=blue!00!black!00,
                      text centered]
\tikzstyle{format1d} = [rounded rectangle,
                      thick,
                      dashed,
                      minimum size=2cm,
					  draw=blue!50!black!80,
					  top color=white,
                      bottom color=blue!00!black!00,                  
                      text centered]
                      
\tikzset{cross/.style={cross out, draw=black, minimum size=2*(#1-\pgflinewidth), inner sep=0pt, outer sep=0pt},
	cross/.default={5pt}}
\usepackage[ normalem]{ulem}
\usepackage{tocloft}

\numberwithin{equation}{section}

\selectfont

\input{Defs.tex}

\begin{document}
\renewcommand{\contentsname}{}
\renewcommand{\refname}{\begin{center}References\end{center}}
\renewcommand{\abstractname}{\begin{center}\footnotesize{\bf Abstract}\end{center}} 

\begin{titlepage}
\ph{preprint}

\vfill

\begin{center}
   \baselineskip=16pt
   {\large \bf Yang-Baxter structure of the extended space
   }
   \vskip 2cm
     Kirill Gubarev$^\dagger{}^*$\footnote{\tt kirill.gubarev@phystech.edu }
     Edvard T. Musaev$^\bullet{}^*$\footnote{\tt emusaev@theor.jinr.ru}
       \vskip .6cm
             \begin{small}
                          {\it
                          $^\dagger$Institute for Information Transmission Problems, 127051, Moscow, Russia\\
                          $^\bullet$Bogoliubov Laboratory of Theoretical Physics, Joint Institute for Nuclear Research, \\ 6, Joliot Curie, 141980 Dubna, Russia\\
                          $^*$Moscow Institute of Physics and Technology, 
                          Laboratory of High Energy Physics, \\
                          9, Institutskii pereulok, 141702, Dolgoprudny, Russia
                          } \\ 
\end{small}
\end{center}

\vfill 
\begin{center} 
\textbf{Abstract}
\end{center} 
\begin{quote}
We construct an analogue of Yang--Baxter deformations defined by a single Killing vector, that is a solution generating transformation in Einstein--Maxwell dilaton theory. We show that these are nothing but a coordinate transformation in a parent theory related to EMd theory by KK reduction. Similarly (almost-abelian) bi-vector Yang--Baxter deformations are coordinate transformations in the doubled space.
\end{quote}

\vfill
\setcounter{footnote}{0}
\end{titlepage}

\tableofcontents

\setcounter{page}{2}

\section{Introduction}

Recently a lot of attention has been drawn to a class of background fields transformations of two-dimensional non-linear sigma-models (the Green--Schwarz superstring) dubbed Yang--Baxter bi-vector deformations. Such a transformation is parameterized by a constant square matrix satisfying the classical Yang-Baxter equation, and preserves Lax representation of field equations and hence classical integrability of the system. The interest to families of integrable sigma-models has long history probably starting with the work \cite{Zakharov:1973pp} where the inverse scattering method has been suggested, and \cite{Cherednik:1981df} where an integrable family of SU(2) $\s$-modes with a deformed Killing form has been presented. Further in \cite{Klimcik:2002zj} this deformation was understood in terms of Yang--Baxter sigma-models, i.e. those deformed by a classical $r$-matrix, and shown to be integrable in \cite{Klimcik:2008eq}. In the context of string theory and supergravity, that is the main interest of the present paper, these models have gained huge interest after the works \cite{Delduc:2013fga,Delduc:2013qra} where the procedure was generalized to sigma-models on coset superspaces and an integrable family of Yang--Baxter deformations of the Metsaev--Tseytlin superstring on $\AdS_5\times \SS^5$ was constructed. To the moment the literature contains a huge amount of results on integrable deformation of $\AdS_n\times \SS^n$ sigma-models which we do not intend to review here instead referring the interested reader to  \cite{Hoare:2021dix,Klimcik:2021bjy,Orlando:2019his, Gubarev:2023jtp,Seibold:2020ouf,Borsato:2022ubq}.

From the supergravity point of view a Yang--Baxter deformation of the superstring sigma-model manifests itself as a redefinition of background fields that is a solution generating transformation in the space of solutions to (generalized) supergravity equations. In particular for the deformations of the $\AdS_5\times \SS^5$  sigma-model found in \cite{Delduc:2013fga} one finds a background that is a solution to a certain generalization of supergravity that includes a Killing vector $I^m$ \cite{Arutyunov:2015mqj,Arutyunov:2015qva}. This appears to be a generalization of the Lunin--Maldacena TsT transformations \cite{Lunin:2005jy} and for the NS-NS sector take the following simple form suggested in \cite{Araujo:2017jkb}
\begin{equation}
\label{eq:deformation0}
    G+B = \big((g+b)^{-1} + \beta\big)^{-1},
\end{equation}
where $g,\,b$ and $G,\,B$ are the initial and final metric and B-field respectively and $\beta$ is a bi-vector $\beta = r^{a b} k_a \wedge k_b$ constructed of the classical r-matrix and Killing vectors of the initial background. Transformation rules for the R-R sector has been found in \cite{Bakhmatov:2017joy}. The field transformation above is non-linear and thus a direct check that it maps a solution to supergravity equations to a solution seems to be nearly impossible and was done perturbatively up to the second order in $\beta$ in \cite{Bakhmatov:2018apn}. The perturbative analysis shows that the sufficient conditions for the deformed background to be a solution are
\begin{equation}
\label{eq:bi-vectorCYBE}
    \begin{aligned}
    \b^{p[m}\partial_p \b^{nk]}&=0, \\
        \partial_m \b^{mn}&=0,
    \end{aligned}
\end{equation}
where the first is the classical Yang--Baxter equation for the matrix $r^{ab}$ given $k_a$ are Killing vectors and the second is the so-called unimodularity condition introduced in \cite{Borsato:2016ose}.

The transformation \eqref{eq:deformation0} becomes linear when formulated in terms of field combinations that transform covariantly under the $\rmO(10,10)$ T-duality symmetry group, that is a $20\times 20$ matrix $\mH$ called the generalized metric
\begin{equation}
\label{eq:genmetric}
    \mH = 
        \begin{bmatrix}
            g - b g^{-1} b & g^{-1}b \\
            - b g^{-1} & g^{-1}
        \end{bmatrix}
\end{equation}
and a function $d = \f - 1/4 \log \det g$ called the invariant dilaton. The deformation is then a linear $\rmO(10,10)$ transformation given by
\begin{equation}
    O_\b = 
        \begin{bmatrix}
            \mathbb{1} & \b \\
            \mathbb{0} & \mathbb{1}
        \end{bmatrix}.
\end{equation}
Using this representation in \cite{Bakhmatov:2018bvp} it was shown that such a bi-vector deformation is a solution generating transformation, and the 11-dimensional analogue dubbed poly-vector deformations has been constructed in \cite{Bakhmatov:2020kul,Gubarev:2020ydf} using the idea of flux covariance of \cite{Borsato:2018idb}. In \cite{Orlando:2019rjg} it has been shown that for a subalgebra of abelian Killing vectors a bi-vector deformation always preserves a given Lax connection.

It is important to mention, that although the transformation is an element of $\rmO(10,10)$ it is not a T-duality transformation, and it is in general not clear why such defined field redefinitions map solutions to solutions and preserve integrability. In other words the question is: what symmetries of the space of string backgrounds do these transformations represent? In this work we approach these questions by considering the simplest form of poly-vector deformations, that are uni-vector deformations. These appear in the standard Kaluza--Klein reduction of a gravitational theory and can be constructed in a complete analogy to deformations by higher rank poly-vectors. From the point of view of the parent higher dimensional theory these are  coordinate transformations of particular form depending linearly on compact coordinate(s). We show that the same is true for bi-vector deformations, that can be understood as coordinate transformations depending linearly on the dual coordinate of the so-called extended space of double field theory. The most fascinating is the origin of the classical Yang--Baxter equation in the form \eqref{eq:bi-vectorCYBE}, that is a consequence of the consistency of the algebra of the coordinate transformations. Certainly, in the uni-vector case this is automatic and no condition appears, that can also be seen from analysis of field equations: any deformation generated by a vector $\alpha = \r^a k_a$ map a solution to a solution given $k_a$ are Killing. Although being pretty trivial from the higher dimensional point of view uni-vector deformations may be of use as solution generating transformations for the Einstein--Maxwell dilaton theory in addition to other approaches known in the literature \cite{Cadoni:2018pav,Vigano:2022hrg,Yazadjiev:2006ew}.

This paper is structured as follows. In Section \ref{sec:univector} we construct uni-vector deformations, provide examples of solutions to Einstein--Maxwell dilaton theory generated by such transformations and show that these are equivalent to coordinate transformation in the parent theory. In Section \ref{sec:bi-vector} we investigate whether bi-vector deformations can be represented as doubled coordinate transformation, show that classical Yang--Baxter equation in the bi-Killing ansatz ensures closure of algebra of such transformations. We explicitly check that the particular class of bi-vector deformations, called almost-abelian, are equivalent to doubled coordinate transformations. In Section \ref{sec:concl} we discuss the results and their connections to integrability properties of 2d sigma-models and possible extensions. Appendix contains technical details expanding calculations in the main text. Calculations based on computer algebra programs Wolfram Mathematica and Cadabra \cite{Peeters:2006kp,Peeters:2018dyg} can be found in Cadabra files on the GitHub repository \cite{Gubarev:2025uni}.

\section{Uni-vector deformations}
\label{sec:univector}

The representation of bi-vector deformations as linear transformations of the generalized metric clearly shows that the deformation parameter should be in some sense a tensor of mixed internal-external components. In the bi-vector case this is achieved by considering a space-time of doubled dimension with coordinates $(x^m, \tilde{x}_m)$, hence $\b^{mn}$ carries one index along the standard coordinate $x^m$ and another along the dual coordinate $\tilde{x}_m$. In other words, the bi-vector stands in the lower left block of the $\rmO(10,10)$ transformation matrix $O_\b$. Further consistency of the full diffeomorphism transformations requires $\beta^{mn}$ to be also a tensor in terms of general coordinate transformations of $x^m$, and the requirement $O_\b \in \rmO(10,10)$ makes it antisymmetric. Postponing the discussion of the bi-vector case to Section \ref{sec:bi-vector}, we will focus here at reproducing the same structures in more conventional terms: General Relativity (Einstein--Maxwell theory) and standard coordinate transformations.

\subsection{Proof of concept: D=4 Kaluza--Klein gravity}\label{sec:proof-of-conc}

To organize a similar deformation tensor in General Relativity we consider its Kaluza--Klein reduction, that splits the metric as
\begin{equation}
\label{eq:4Dint}
    ds^2 = e^\f g_{mn} dx^m dx^n + e^{-\f}\big(dz + \mA_m dx^m\big)^2.
\end{equation}
and impose no dependence of the fields on $z$. For simplicity in this section we consider the parent theory to be four-dimensional, i.e. $m,n,\dots = 1,2,3$, the general case will be covered in the Section \ref{sec:emd}. Note also, that $z$ might chosen to be the time-like direction. The action is taken to be in the standard Einstein--Hilbert form
\begin{equation}\label{GL4actionInv}
    S_{GL(4)} = \int d^4x \sqrt{-G} R[G],
\end{equation}
and we omit any discussion of the boundary effects, hence there is no Gibbons--Hawking term\footnote{It is worth to mention though, that these might be important since i) the formalism includes integration by parts, ii) in examples with a horizon the latter deforms and doubles.}. The dimensionally reduced action then reads 
\begin{equation}\label{GL4action}
    S_{3} = \int d^3x \sqrt{-g} \Big( R[g] - \frac12 \partial_{m} \phi \, \partial^{m} \phi - \frac14 e^{- 2 \phi} \mF^{m n} \mF_{m n} \Big),
\end{equation}
where $R[g]$ is the usual Ricci tensor for the 3-dimensional metric $g_{m n}$, and $\mF_{m n} = 2 \partial_{[m} \mA_{n]}$. For further discussion it is important to mention some similarities with double field theory. If one was to follow the logic of double field theory one would start with the action \eqref{GL4action} (the analogue of Type II supergravity here) and realize that it is secretly invariant under a larger group of symmetries, that is 4-dimensional diffeomorphisms. The action fully covariant under these symmetries would be \eqref{GL4actionInv} (the analogue of the double field theory action of \cite{Hohm:2010pp}). 

In the component form the metric in the split \eqref{eq:4Dint} reads 
\begin{equation}
\begin{aligned}
        G_{M N} = 
        \begin{bmatrix}
		       	e^{\phi} g_{m n} + e^{-\phi} \mA_{m} \mA_{n} &  e^{- \phi} \mA_{m} \\
		    	e^{- \phi} \mA_{n} &   e^{- \phi}
	    \end{bmatrix},
\end{aligned}
\end{equation}
that is similar to the form of generalized metric \eqref{eq:genmetric}. The vielbein defined as usual as $G_{M N} = E^{A}{}_{M} E^{B}{}_{N} G_{A B}$ can be written in the upper-triangular parametrisation as follows
\begin{equation}\label{GL4vielbeins}
\begin{aligned}
        E^{A}{}_{M} = 
        \begin{bmatrix}
		       	e^{\frac\phi2} e^{a}{}_m & e^{- \frac\phi2} \mA_{m} \\
		    	0 &  e^{- \frac\phi2}
	     \end{bmatrix}, \quad
	     E_{A}{}^{M} = 
        \begin{bmatrix}
		       	e^{- \frac\phi2} e_{a}{}^m & - e^{- \frac\phi2} \mA_{a} \\
		    	0 &  e^{\frac\phi2}
	     \end{bmatrix}, \quad \partial_{M} = (\partial_{m}; \partial_{4} \equiv 0),
\end{aligned}
\end{equation}
where $G_{AB}$ is taken to be the flat Minkowski metric. Since $G \in \rmGL(4)$ a uni-vector deformation must be also an element of $\rmGL(4)$. To arrive at the explicit form of the deformation and to keep the analogy let us follow the logic adopted for the construction of poly-vector deformations in \cite{Gubarev:2020ydf}. For that consider breaking of the full $\rmGL(4)$ group under the Kaluza--Klein split, that decomposes the generators $T_M{}^N$ intro three sets
\begin{equation}
    \mathrm{bas}\,{\mathfrak{gl}(4)} = \{T_m{}^4, T_m{}^n, T_4{}^n\}.
\end{equation}
The elements $T_m{}^n$ generate $\rmGL(3)$ transformations, the elements $T_4{}^m$ generate shifts of the vector $\mc{A}_m$, and the remaining elements $T_m{}^4$ are the ones that generate deformations. Explicitly we have
\begin{equation}
\label{eq:monovecdef}
    O_\a = \exp \big(\a^m T_m{}^4\big) = 
        \begin{bmatrix}
            \mathbb{1} & \mathbb{0} \\
            \a^m & 1
        \end{bmatrix},
\end{equation}
where $\a^m = \a^m(x)$ denote components of a vector field, and the deformed metric becomes
\begin{equation}
\label{eq:Gdeform}
\begin{aligned}
        \tilde{G}_{M N} & =  
        \begin{bmatrix}
		       	e^{\phi} g_{m n} + e^{- \phi} \mA_{m} \mA_{n} & e^{\phi} \a_{m} + e^{- \phi} \mA_{m} (1 + \mA_{k} \a^{k}) \\
		    	e^{\phi} \a_{n} + e^{- \phi} \mA_{n} (1 + \mA_{k} \a^{k}) &  e^{\phi} \a_{k} \a^{k} + e^{- \phi} (1 + \mA_{k} \a^{k})^2
	    \end{bmatrix}, \\[1em]
        &= \begin{bmatrix}
		       	e^{\tilde \phi} \tilde g_{m n} + e^{-\tilde \phi} \tilde \mA_{m} \tilde \mA_{n} & e^{- \tilde \phi} \tilde \mA_{m} \\
		    	e^{- \tilde \phi} \tilde \mA_{n} &  e^{- \tilde \phi}
	    \end{bmatrix}.
\end{aligned}
\end{equation}
Here the first line represents explicit action of the deformation and the second allows to recover deformed 3d metric, gauge field and the dilaton. Indices on the RHS are raised and lowered by the initial (not deformed) metric making the transformation written in terms of lower dimensional fields highly non-linear:
\begin{equation}
    \label{eq:deformedcomponents}
    \begin{aligned} 
        e^{- \tilde \phi} & =  e^{\phi} \a_{k} \a^{k} + e^{- \phi} (1 + \mA_{k} \a^{k})^2 \,, \\
        \tilde \mA_{m} & =  e^{\tilde \phi} (e^{\phi} \a_{m} + e^{- \phi} \mA_{m} (1 + \mA_{k} \a^{k})) \,, \\
        \tilde g_{m n} & =  e^{- \tilde \phi} (e^{\phi} g_{m n} + e^{- \phi} \mA_{m} \mA_{n}) - e^{- 2 \tilde \phi} \tilde \mA_{m} \tilde \mA_{n}\,. 
    \end{aligned}
\end{equation}
This is an analogue of open/closed map \cite{Seiberg:1999vs} on the case of uni-vector. It is not clear whether this transformation holds a similar meaning, and the authors are not aware of relevant considerations in the literature.

To derive conditions for such defined deformation to respect equations of motion it is most convenient to formulate the latter in terms of anholonomicity coefficients. These are defined as
\begin{equation}\label{GL4fluxes}
\begin{aligned}
\mF_{A B}{}^{C} = -2 E_{A}{}^{M} E_{B}{}^{N} \partial_{[M} E_{N]}{}^{C},
\end{aligned}
\end{equation}
and are not necessarily constant in general. In what follows these will be referred to as fluxes to match the terminology of supergravity and the exceptional field theory literature, where generalized anholonomicity coefficients contain literally fluxes of gauge fields as components. Since the fluxes $\mF_{AB}{}^C$ have only flat indices $A, B, C$ on which the deformation $(O_\a)_M{}^N$ cannot act linearly, and $\det O_\a = 1$, it is natural to require them to be invariant. As a bonus this will also ensure invariance of equations of motion. Indeed the higher dimensional Ricci scalar can be written completely in terms of the fluxes as (see details in Appendix \ref{appRicciGTR})
\begin{equation}\label{eq:ricciscalarGL4}
\begin{aligned}
R[G] & =  - \frac{1}{2}\, {\mF}_{A B}\,^{C} {\mF}_{C D}\,^{A} {G}^{B D} - \frac{1}{4}\, {\mF}_{A B}\,^{C} {\mF}_{D F}\,^{G} {G}_{C G} {G}^{A D} {G}^{B F} \\
&\quad\, - {\mF}_{A} {\mF}_{B} {G}^{A B} + 2 \, {\partial}_{M}{{\mF}_{A}}\,  {E}^{A}\,_{N} {G}^{M N},
\end{aligned}
\end{equation}
where $\mF_{A} = \mF_{A B}{}^{B} = - E^{-1} \dt_{M} (E \, E_{A}{}^{M})$ and $E=\det E_M{}^A$. Next, when substituted into the action the last term can be integrated by parts and rendered in the following form
\begin{equation}
 2 \, \sqrt{-G} {\partial}_{M}{{\mF}_{A}}\,  {E}^{A}\,_{N} {G}^{M N} =  2 \, {\partial}_{M} (\sqrt{-G} {{\mF}_{A}}\,  {E}_{B}{}^{M} {G}^{B A} ) + 2 \, \sqrt{-G} \mF_{A} \mF_{B} G^{A B},
\end{equation}
implying that the full action is quadratic in fluxes and does not include derivative terms. Hence, invariance of fluxes under a uni-vector deformation leaves value of the action unchanged and keeps it at the same minimum value as prior to the deformation. Therefore we require
\begin{equation}
\begin{aligned}
        \mF_{A B}{}^{C} = {\mF}'_{A B}{}^{C}.
\end{aligned}
\end{equation}

Important is the issue of the upper trinagular gauge fixing, that is spoiled by the deformation \eqref{eq:monovecdef} and the deformed vielbein becomes 
\begin{equation}\label{GL4deformation}
\begin{aligned}
        E'^{A}{}_{M} = O_{M}{}^{N} E^{A}{}_{N} = 
        \begin{bmatrix}
		       	e^{\frac\phi2} e^{a}{}_m & e^{- \frac\phi2} \mA_{m} \\
		    	e^{\frac\phi2} \a^{a} &  e^{- \frac\phi2} (1 + \mA_{k} \a^{k})
	      \end{bmatrix}.
\end{aligned}
\end{equation}
To stay in the ``gravity frame'', i.e. to have a theory symmetric under 3d diffeomorphisms and gauge transformations $\d \mc{A} = d\l$, one should additionally perform a local $\rmSO(4)$ transformation. Certainly, the choice of the gauge does not affect the metric, both deformed and undeformed, and as we have seen above, the action is written completely in terms of fluxes and does not include the vielbein explicitly. Since the action is also invariant under local Lorentz transformations its value stays in the same minimum after going back to the upper triangular gauge. We conclude that the condition of the flux invariance is sufficient for a transformation to preserve equations of motion. Explicit calculation (see \texttt{GL4-flux-deformations} in {\color{red} \cite{Gubarev:2025uni}}) shows that a sufficient condition for that is 
\begin{equation}
\begin{aligned}
        \mL_{\a} \, \phi = 0, \quad \mL_{\a} \, e^{a}{}_{m} = 0, \quad \mL_{\a} \, \mA_{m} = 0,
\end{aligned}
\end{equation}
i.e. $\a^m$ is a Killing vector for the initial background. It is important to mention here that one should rather require $\mc{L}_\a g_{mn}=0$ instead of the same condition for the vielbein, since these are not always equivalent. For a single Killing vector one can always choose the frame such that both Lie derivatives of the metric and the vielbein vanish. This no longer holds for bi-vector deformations, where one needs at least two Killing vectors. In an upcoming work we show, that while fluxes do change in general, both for uni-vector and bi-vector deformation the action transforms by a full derivative and hence equations of motion still hold. Given that, we will continue using the condition $\mL_{\a} \, e^{a}{}_{m} = 0$ in what follows.

To gain more information about the geometric meaning of the uni-vector field transformation introduced above it is suggestive to consider a simple example. The form of the deformed metric \eqref{eq:Gdeform} suggests that uni-vector deformations allow to generate new solutions to pure Einstein equations, which is however not completely true. To see that consider the metric of the Schwarzschild black hole
\begin{equation}
    ds^2 = - f(r)^{-1}dt^2 + f(r) dr^2 + r^2 d\q^2 + r^2 \sin^2 \q d\vf^2,
\end{equation}
where $f(r)^{-1} = 1 - \fr{2M}{r}$. This space-time has three rotational Killing vectors and one Killing vector along $t$. Since one can always align axes such that the rotational Killing vector is along the coordinate $\phi$ in the metric above, we take
\begin{equation}
    \a = \g \, \dt_\vf,
\end{equation}
and identify the $z$ coordinate of the formalism with $t$. After applying the rules \eqref{eq:deformedcomponents} we arrive at the following static metric
\begin{equation}
    ds^2 = -dt^2 \left(1 - \fr{2M}{r} - \g^2 r^2 \sin^2 \q \right) + 2\g r^2 \sin^2 \q  dt d\vf +\fr{1}{1-\fr{2M}{r}} dr^2 + r^2 d\W^2,
\end{equation}
where $d\W^2 = d\q^2 + \sin^2\q d\vf^2$ is the standard spherical angle element. Explicit check shows that it solves Einstein equations as it should do. It is easy to see that upon the following change of coordinates 
\begin{equation}
    \vf' = \vf + \frac{1}{2} \g t
\end{equation}
this turns into the initial Schwarzschild black hole. 

In this case a uni-vector deformation is simply a coordinate transformation in extended (4d) space, that as we show in Section \ref{sec:monocoord} will be true in general, although its explicit form is highly non-trivial. We will arrive at the same conclusions in Section \ref{sec:bi-vector} for a special class of bi-vector deformations of supergravity solutions. This statement is non-trivial as the coordinate transformation acts in a specially constructed extended space and generated solutions are indeed new from the point of view of Einstein--Maxwell dilaton theory or of supergravity.

\subsection{A general approach: Einstein--Maxwell dilaton gravity}\label{sec:emd}

Let us now consider the case of general dimensions and allow for non-vanishing cosmological constant, that gives access to (conformal) AdS solutions. The action is given by the following expression
\begin{equation}\label{GLDaction}
    \begin{aligned}
    S_{EMD} & = \int d^Dx \sqrt{|g|} \Big( R[g] - \frac{1}{2} \partial_{m} \phi \, \partial^{m} \phi \mp \frac{1}{4} e^{- 2 a \phi} \mF^{m n} \mF_{m n} - 2 \Lambda e^{(D\gamma + \beta)\phi}\Big) \\
     &- 2 \gamma \int d^Dx \dt_{m} (\sqrt{|g|} g^{m n} \dt_{n} \phi\Big),
     \end{aligned}
\end{equation}
where $R[g]$ is the ordinary Ricci tensor for the $D$-dimensional metric $g_{m n}$, $\mF_{m n} = 2 \partial_{[m} \mA_{n]} $, and $\Lambda$ is the cosmological term. The upper sign corresponds to the case when $g_{m n}$ has Minkowski signature and the lower one when it has Euclidean signature. In the latter case we will assume that the additional (Kaluza--Klein) coordinate is the time, $\g$ and $\b$ are constants to be tuned later. To relate our notations to those of \cite{Pope:2024ncb,Cvetic:2014vsa} one replaces $\phi \to 2 \phi$, $\mA_{m} \to 2 \mA_{m}$ and $a \to \frac{a}{2}$. In the case $D=4$ the  action in the form above can be electromagnetically dualized using
\begin{equation}
    \bar{\mF} = e^{-2a\phi} * \mF
\end{equation}
into the action (dropping full derivative term)
\begin{equation}\label{GLDactiondual}
    S_{\text{dual}} = \int d^4x \sqrt{|g|} \Big( R[g] - \frac{1}{2} \partial_{m} \phi \, \partial^{m} \phi \mp \frac{1}{4}e^{2 a \phi} \bar{\mF}^{m n} \bar{\mF}_{m n} - 2 \Lambda e^{(4\gamma + \beta)\phi}\Big),
\end{equation}
used in \cite{Pope:2024ncb} to investigate stability of electrically charged EMD black holes. 

For this theory to be a reduction of a $(D+1)$-dimensional General Relativity by the standard Kaluza--Klein ansatz the constants must be chosen as follows
\begin{equation}
    a = \frac{(D-1)}{\sqrt{2(D-1)(D-2)}}, \quad \gamma = \frac{1}{\sqrt{2(D-1)(D-2)}}, \quad \beta = - \sqrt{\frac{(D-2)}{2(D-1)}}.
\end{equation}
As before, this implies that the action \eqref{GLDaction} is secretly invariant under the full diffeomorphism symmetry in $D+1$ dimensions and hence can be written as
\begin{equation}\label{GLDactionInv}
    S_{GL(D+1)} = \int d^{D+1}x \left(\sqrt{|G|} R[G] - 2 \Lambda \sqrt{|G|}\right),
\end{equation}
where the higher dimensional metric  $G_{MN}$ encodes the metric, gauge and scalar fields in lower dimensions upon the standard Kaluza--Klein ansatz
\begin{equation}
    \label{eq:highermetric}
    ds^2 = e^{2 \gamma \f} g_{mn} dx^m dx^n \pm e^{2\beta\f}\big(dz + A_m dx^m\big).
\end{equation}
The expression \eqref{eq:ricciscalarGL4} for the Ricci scalar in terms of anholonomic coefficients (fluxes) does not depend on the dimensions. As before the vielbein corresponding to the higher dimensional metric can be written in the upper-triangular parametrization as follows
\begin{equation}\label{GLDvielbeins}
\begin{aligned}
        E^{A}{}_{M} = 
        \begin{bmatrix}
		       	e^{\gamma\phi} e^{a}{}_m & e^{\beta\phi} \mA_{m} \\
		    	0 &  e^{\beta\phi}
	     \end{bmatrix}, \quad
	     E_{A}{}^{M} = 
        \begin{bmatrix}
		       	e^{- \gamma\phi} e_{a}{}^m & 0 \\
		    	- e^{- \gamma\phi} \mA_{a} &  e^{-\beta\phi}
	     \end{bmatrix},
\end{aligned}
\end{equation}
and the determinant reads $|G| = |\det G_{MN}| = e^{2(D\gamma + \beta)\phi} |\det g_{mn}| = e^{2(D\gamma + \beta)\phi} |g|$. The flat metric is taken to be $G_{AB}=\mathrm{diag}[\h_{ab},\pm 1]$. On principle, the block $\h_{ab}$ may be of the Lorenzian signature and hence the higher-dimensional space-time will have two time-like directions. However, we will always consider examples where the time is either among $x^m$ or identified with $x^{D+1}$.

In such a parametrization the non-vanishing components of fluxes take the following form
\begin{equation}
    \begin{aligned}
    \mF_{a \, D+1}{}^{D+1} & = - \beta e^{-\gamma \phi} e_{a}{}^{m} \partial_{m}{\phi}, \\
    \mF_{a b}{}^{D+1} & = - e^{(-2\gamma + \beta) \phi} e_{a}{}^{m} e_{b}{}^{n} \mF_{m n}, \\
    \mF_{a b}{}^{c} & = e^{-\gamma \phi} f_{a b}{}^{c} - 2 \gamma e^{-\gamma \phi} e_{[a}{}^{m} \delta_{b]}{}^{c} \partial_{m}{\phi} ,
    \end{aligned}
\end{equation}
where the lower dimensional anholonomic coefficients are defined accordingly
\begin{equation}
    f_{ab}{}^{c} = -2 e_{a}{}^{m} e_{b}{}^{n} \partial_{[m}e_{n]}{}^{c}.
\end{equation}
Equations of motion for the action \eqref{GLDactionInv}  can be written completely in terms of the fluxes and their (first) derivatives, and in the upper triangular parametrization are equivalent to the field equations \eqref{GLDaction}. The latter have the following form
\begin{equation}
\label{glDequations1}
\begin{aligned}
    R_{m n} - \frac{1}{2} R g_{m n} + \L e^{(D\gamma + \beta)\phi} g_{m n} &= T_{mn} \\ 
    \nabla_{m} ( e^{- 2 a \phi} \mF^{m n} ) &= 0\,, \\
    \Box \phi \pm \frac{a}{2} e^{- 2 a \phi} \mF_{m n} \mF^{m n}  & = 2 (D\gamma + \beta) \Lambda e^{(D\gamma + \beta)\phi}\,,    
\end{aligned}
\end{equation}
where
\begin{equation}
    T_{mn}  =\frac{1}{2} \left( \dt_{m} \phi \, \dt_{n} \phi - \frac{1}{2} (\dt \phi)^2 g_{m n} \right)  \pm \frac{1}{2} e^{-2 a \phi} \left( \mF_{m k} \mF_{n}{}^{k} - \frac{1}{4} \mF_{k l} \mF^{k l} g_{m n} \right).
\end{equation}

Following the discussion in the previous section we now  consider a uni-vector deformation of the fields, written in the form
\begin{equation}\label{GLDdeformation}
\begin{aligned}
        E'^{A}{}_{M} = O_{M}{}^{N} E^{A}{}_{N} = 
        \begin{bmatrix}
		       	e^{\gamma\phi} e^{a}{}_m & e^{\beta\phi} \mA_{m} \\
		    	e^{\gamma\phi} \a^{a} &  e^{\beta\phi} (1 + \mA_{k} \a^{k})
	      \end{bmatrix}, \quad
        O_{M}{}^{N} = 
        \begin{pmatrix}
                \d_m{}^n && 0 \\
                \\
                \a^{n} && 1
        \end{pmatrix}.
\end{aligned}
\end{equation}
Such a transformation preserves determinant of the metric $G_{(D+1)}$, therefore the condition that the fluxes do not transform
\begin{equation}
\begin{aligned}
        \mF_{A B}{}^{C} = {\mF}'_{A B}{}^{C},
\end{aligned}
\end{equation}
will map a solution into a solution. In other words, the minimum of the action does not change. The deformed metric can be written explicitly as follows
\begin{equation}
\begin{aligned}
        \tilde{G}_{M N} &=
        \begin{bmatrix}
		       	e^{2 \gamma \phi} g_{m n} \pm e^{2 \beta \phi} \mA_{m} \mA_{n} & e^{2\gamma\phi} \a_{m} \pm e^{2\beta\phi} \mA_{m} (1 + \mA_{k} \a^{k}) \\
		    	e^{2\gamma\phi} \a_{n} \pm e^{2\beta\phi} \mA_{n} (1 + \mA_{k} \a^{k}) &  e^{2\gamma\phi} \a_{k} \a^{k} \pm e^{2\beta\phi} (1 + \mA_{k} \a^{k})^2
	    \end{bmatrix},
\end{aligned}
\end{equation}
and implies the following  transformation rules for the lower dimensional fields
\begin{equation}
\begin{aligned}
\label{glDdeformedsolution1}
e^{2\beta\tilde \phi} & =  \pm e^{2\gamma\phi} \a_{k} \a^{k} + e^{2\beta\phi} (1 + \mA_{k} \a^{k})^2 \,, \\
\tilde \mA_{m} & =  \pm e^{-2\beta\tilde \phi} (e^{2\gamma\phi} \a_{m} \pm e^{2\beta\phi} \mA_{m} (1 + \mA_{k} \a^{k})) \,, \\
\tilde g_{m n} & =  e^{-2\gamma\tilde \phi} (e^{2 \gamma \phi} g_{m n} \pm e^{2 \beta \phi} \mA_{m} \mA_{n}) \mp e^{2(\beta-\gamma)\tilde \phi} \tilde \mA_{m} \tilde \mA_{n}\,. 
\end{aligned}
\end{equation}
These can be nicely written in terms of the blocks of the full metric $\tilde{G}_{MN}$
\begin{equation}
\label{eq:block_transformD}
    \begin{aligned}
        \tilde{G}_{mn} & = G_{mn}, \\
        \tilde{G}_{mz} & =G_{mz} + G_{mn}\a^n,\\
        \tilde{G}_{zz} & = G_{zz} + 2 G_{z m} \a^m + G_{mn}\a^m \a^n.
    \end{aligned}
\end{equation}
The condition that the fluxes are invariant under the transformation, and hence a solution is mapped to a solution, restricts the vector  $\a^{m}$ to be a Killing vector of the initial background (see \texttt{GL4-flux-deformations} in \cite{Gubarev:2025uni})
\begin{equation}
\begin{aligned}
        \mL_{\a} \, \phi = 0, \quad \mL_{\a} \, e^{a}{}_{m} = 0, \quad \mL_{\a} \, \mA_{m} = 0.
\end{aligned}
\end{equation}

From the explicit form of the transformation one concludes that the field transformation as defined above certainly produces new solutions in the $D$-dimensional Einstein--Maxwell dilaton theory, that we illustrate by several examples below. However, as we prove below in Section \ref{sec:monocoord} the transformation rules \eqref{glDdeformedsolution1} are nothing but a coordinate transformation in $D+1$ dimensions, given the vector $\a =\a^m\dt_m$ is Killing and does not depend on $x^{D+1}$. This gives a strong hint on a similar understanding of bi-vector Yang--Baxter deformations.

\subsection{More examples}

Let us now illustrate the above construction by examples of solution generation in Einstein--Maxwell dilaton theory with and without the cosmological term.

\subsubsection{Vanishing cosmological constant, \texorpdfstring{$\L = 0$}{Lambda=0}}

1. \textbf{Flat D=3 space,} that corresponds to $\gamma = \frac{1}{2}$, $\beta = - \frac{1}{2}$ and $a = 1$
    \begin{equation}\label{flat3d}
        ds^2 = - dt^2 + dr^2 + r^{2} d \theta^2, \quad \phi = 0, \quad \mA_{m} = 0.
    \end{equation}
The Killing vector is taken to be $\a = \h \, \partial_{\q}$ and the deformed fields read
\begin{equation}\label{flat3dDeformed}
    \begin{aligned}
       ds^2 & = (1+\h^2 r^2)\big(-dt^2 + dr^2\big) + r^2 d\q^2,\\
       \tilde{\mc{A}} & = \frac{\h r^2}{\h^2 r^2+1} d\q, \\
       \tilde{\phi}& = -\ln \left(1 + \h^2 r^2\right).
    \end{aligned}
\end{equation}
This is a solution to  \eqref{glDequations1} that cannot be expressed as a combination of 3d diffeomorphisms and gauge transformations of the initial solution \eqref{flat3d} since the invariant quantities, such as the Ricci scalar and the field strength for $\mA_{m}$, change: from vanishing values $R = 0$, $\mF_{mn} = 0$ to (\ref{flat3dDeformed})
\begin{equation}\label{Rdefflat3d}
   \tilde{R} = \frac{2 \h^2 \left(\h^2 r^2-1\right)}{\left(\h^2 r^2+1\right)^3},
        \quad \tilde{\mF}  = \frac{2 \h r}{\left(\h^2 r^2+1\right)^2} dr \wedge d\q.
\end{equation}
Therefore from the 3D point of view the deformation is nontrivial.

2. \textbf{Flat D=3 space, a different gauge.} One of the subtle points in the poly-vector deformation procedure is that the result depends on the gauge chosen for the initial gauge fields. This is expected since the gauge potential rather than its field strength enters the deformation. To see that explicitly let us consider the same flat 3D space-time, however with a constant non-vanishing gauge field 
    \begin{equation}\label{flat3d2}
        ds^2 = - dt^2 + dr^2 + r^{2} d \theta^2, \quad \phi = 0, \quad \mA = h_{t}dt +h_{r} dr + h_{\theta} d \q.
    \end{equation}
Here $h_m$ are arbitrary constants. The deformation along the same  Killing vector $\a = \h \, \partial_{\varphi}$ reads
\begin{equation}\label{flatdiffgauge}
    \begin{aligned}
        ds^2 &= -(\h^2 r^2+\k^2) dt^2  + (\h^2 r^2+\k^2)dr^2
        +r^2 (\h h_t dt + \h h_r dr  - d\q)^2 \\
        \tilde{\mA} &= \frac{1}{\h^2 r^2+\k^2}\left(h_{t} \k dt + h_{r}  \k  dr + (\h h_{\theta}^2+\h r^2+h_{\theta})d\q\right),\\
        \tilde{\phi}& = -\ln \left(\h^2 r^2+\k^2\right)
    \end{aligned}
\end{equation}
where we denote $\k = \h h_\q + 1$ for clarity of notations. Explicit form of the Ricci tensor
\begin{equation}
    \tilde{R} = \frac{2 \h^2 \left(\h^2 r^2-
    \k^2\right) }{\left(\h^2 r^2+\k^2\right){}^3},
\end{equation}
shows that this is yet another family of solutions parametrised by the deformation parameter $\h$ and by constant values $h_t,h_r,h_\q$ of the initial gauge field components.

Given that such deformations are simply coordinate transformations in the higher dimensional theory, that include lower dimensional gauge transformations, it is of no surprise that diffeomorphisms performed in different order give different field components. Certainly, the full 4-dimensional Ricci scalar does not change.

3. \textbf{Black hole in 4D.} Now we turn to four dimensions, that means $\gamma = \frac{1}{2\sqrt{3}}$, $\beta = -\frac{1}{\sqrt{3}}$, $a = \frac{\sqrt{3}}{2}$, and consider the Schwarzschild black hole solution
\begin{equation}
    ds^2 = - f(r) dt^2 + f(r)^{-1} dr^2 + r^2 (d\theta^2 + \sin^2\theta d\varphi^2), \quad \phi = 0, \quad \mA_{m} = 0,
\end{equation}
where $f(r)=1 + \frac{r_{g}}{r}$. This is the same solution taken as an example in the end of Section \ref{sec:proof-of-conc}, however now we embed it into the (KK reduced) 5D gravity theory or equivalently into the Einstein--Maxwell dilaton theory. The Killing vector is the same though
\begin{equation}
    \a = c \, \partial_{\varphi}.
\end{equation}
The deformed background will be
\begin{equation}
    \begin{aligned}
        ds^2 & = \sqrt{1+c^2 r^2 \sin ^2\theta } \left[- f(r) dt^2 + f(r)^{-1} dr^2 + r^2 \Big(d\theta^2 + \fr{\sin^2\theta}{1+c^2 r^2 \sin ^2\theta }d\varphi^2\Big)\right], \\
        \mc{A} & = \frac{c\, r^2 \sin^2 \q  }{1 + c^2 r^2 \sin^2 \q } d\varphi,\\
        \tilde{\phi}& = -\frac{\sqrt{3}}{2}  \ln \left(1 + c^2 r^2 \sin ^2\theta \right),
    \end{aligned}
\end{equation}
which solves (\ref{glDequations1}) (\texttt{Gibbons\_Maeda\_EDM\_solution\_Rm=0.nb} in \cite{Gubarev:2025uni}). The deformation is again non-trival since the Ricci flat space-time with vanishing electromagnetic field strength turns into 
\begin{equation}\label{BHdef1}
    \begin{aligned}
    \tilde{R} & = \frac{3 c^4 r \sin ^2\theta  \left(r- r_g \sin ^2\theta \right)}{2 \left(1 + c^2 r^2 \sin ^2 \theta \right)^{5/2}}, \\
    \mc{F} & = -\frac{2 c r \sin ^2\theta }{\left(1+c^2 r^2 \sin ^2\theta \right)^2} dr\wedge d\phi-\frac{2 c r^2  \sin \theta  \cos \theta }{\left(1+c^2 r^2 \sin ^2\theta \right)^2} d\q \wedge d\phi 
    \end{aligned}
\end{equation}
If only real coordinate transformations and real deformations are allowed this is not equivalent to the Gibbons--Maeda black hole solution, whose explicit form is given below. To our knowledge, this is a new solution to the Einstein--Maxwell dilaton gravity equations.

4. \textbf{D=4 Gibbons--Maeda black hole.} Let us now perform a deformation of the Gibbon--Maeda black hole solution found in \cite{GIBBONS1988741}, that is $\gamma = \frac{1}{2\sqrt{3}}$, $\beta = -\frac{1}{\sqrt{3}}$. This solves equations of the Einstein--Maxwell dilaton gravity for arbitrary $a$, i.e. existence of the higher dimensional GR parent theory is not required. Explicitly the solution reads
\begin{equation}
    \begin{aligned}
        ds^2 &= - \Delta dt^2 + \Delta^{-1} dr^2 + R^2 (d\theta^2 + \sin^2(\theta)d\varphi^2),\\
        e^{-4a\phi} &= F_{-}^{\frac{4a^2}{1+4a^2}}, \quad \mA = \frac{1}{2} q \cos(\theta) d\varphi,\\
        \Delta &= F_{+} F_{-}^{\frac{1-4a^2}{1+4a^2}}, \quad R^2 = r^2 F_{-}^{\frac{8a^2}{1+4a^2}},\\
        F_{\pm} &= 1 \pm \frac{r_{\pm}}{r}, \quad q = \sqrt{\frac{r_{+}r_{-}}{1+4a^2}},
    \end{aligned}
\end{equation}
where $q$ is the magnetic charge of the black hole and $r_{+}$, $r_{-}$ are its horizons. Existence of two horizons is precisely the reason why the deformation found in the previous example is not related to this solution by a real coordinate transformation for real deformation parameter $c$.

Let us now for simplicity fix    $a = \frac{\sqrt{3}}{2}$ and consider the same deformation along $        \a = c \, \partial_{\varphi}$. The deformed solution can be written as follows
\begin{equation}
    \begin{aligned}
        ds^2 & = -\frac{(r-r_{+}) U}{r-r_{-}} dt^2  + 
        \frac{r U}{r-r_{+}} dr^2 + r (r-r_{-}) U d\q^2 + \frac{\sin ^2(\theta ) (r-r_{-})^2}{U}  d\varphi^2, \\
        \tilde{\mc{A}} & = \frac{c\,r^2 \sin ^2(\theta )+\cos (\theta ) \left(c\,r_{-} r_{+} \cos (\theta )+\sqrt{r_{-} r_{+}}\right)}{c(c\,r^2 \sin ^2(\theta )+c \,r_{-} r_{+} \cos ^2(\theta )+2 \cos (\theta ) \sqrt{r_{-} r_{+}})+1}d\varphi,\\
        \tilde{\phi} & = -\sqrt{3} \ln \left(U\right).
    \end{aligned}
\end{equation}
where  
\begin{equation}
   U^2=\frac{(r-r_{-}) \left(c(c\,r^2 \sin ^2(\theta )+c \,r_{-} r_{+} \cos ^2(\theta )+2 \cos (\theta ) \sqrt{r_{-} r_{+}})+1\right)}{r}
\end{equation}
As before, this is not related to the initial solution by a coordinate and gauge transformations.

5. \textbf{Flat D=3 space.} As a final example we consider less trivial  deformation of the flat space (\ref{flat3d}), that is when the Killing vector depends on coordinates. We take the Killing vector corresponding to boosts along the $x$ coordinate
\begin{equation}
    \begin{aligned}
        \alpha & = c \, (x \partial_{t} + t \partial_{x}).
    \end{aligned}
\end{equation}
Deformed fields are given by
\begin{equation}
    \begin{aligned}
        d\tilde{s}^2 & = -dt^2 + dx^2 + dy^2 -c^2 \Big(\big(tdt - xdx\big)^2 + \big(x^2 - t^2\big)dy^2\Big), \\
        \tilde{\mA} & = c\fr{-x dt + t dx}{1+ c^2(t^2 - x^2)}, \\
        e^{-\f} & = 1 + c^2(t^2 - x^2). 
    \end{aligned}
\end{equation}
This background solves the equations of motion and gives non-trivial $\tilde{R}$ and $\tilde{F}_{m n}$ (\texttt{Flat\_3d\_solution\_boost.nb} in \cite{Gubarev:2025uni}):
\begin{equation}
    \begin{aligned}
        \tilde{R} & = 2 c^2 \fr{1 - c^2 (t^2 - x^2)}{1 + c^2 (t^2 - x^2)},\\
        \tilde{\mF} & = -\fr{2 c}{1 + c^2 (t^2 - x^2)}dt \wedge dx,
    \end{aligned}
\end{equation}
Note that the deformed space-time has a singularity both in the Ricci curvature scalar and the field strength tensor. This is a common feature of all polyvector deformations, when Killing vectors are taken not along a compact isometry direction. Unfortunately, at this moment we are not able to comment on how one should treat this kind of behavior. 

\subsubsection{Non-vanishing cosmological constant, \texorpdfstring{$\L \neq 0$ (conformal Ad$\mathbb{S}_{4}$)}{Lambda <> 0}}

If the cosmological constant does not vanish and is negative the field equations admit a solution of the anti de Sitter geometry. To be concrete let us consider the particular case of D=4 that is $\gamma = \frac{1}{2\sqrt{3}}$, $\beta = -\frac{1}{\sqrt{3}}$, $a = \frac{\sqrt{3}}{2}$ and denote $\Lambda = - \frac{6}{R^2}$. 
In this case the parent 5D theory admits an $\AdS_5$ solution of the form
\begin{equation}
    ds_{\text{Ad}\mathbb{S}_{5}}^2 = \frac{R^2}{x_4^2} (-dx_1^2 + dx_2^2 + dx_3^2 + dx_4^2 + dx_5^2),
\end{equation}
that upon dropping the coordinate $x_5$ gives the following 4D solution to Einstein--Maxwell dilaton equations
\begin{equation}\label{confAdS4solution}
    ds_{\text{conf Ad}\mathbb{S}_{4}}^2 = \frac{R^2}{x_4^3} (-dx_1^2 + dx_2^2 + dx_3^2 + dx_4^2), \quad \mA_{m} = 0, \quad \phi = \sqrt{3}\ln{x_4}. 
\end{equation}
The Ricci scalar becomes $R=-\frac{45 \, x_4}{2 R^2}$. Its deformation along the Killing vector
\begin{equation}
    \alpha = c_1 \dt_1 + c_2 \dt_2 + c_3 \dt_3,
\end{equation}
gives the following background
\begin{equation}
    \begin{aligned}
        ds^2 & = R^2 \big(1+c^2 R^2\big)^{\fr12} ds_{c\AdS}^2-R^4 \big(1+c^2 R^2\big)^{-\fr12}\frac{(c \cdot dx )^2}{x_4^3},\\
        \mc{A} & = R^2\big(1+c^2 R^2\big)^{-\fr12} (c \cdot dx), \\
        \tilde{\phi}& = -\frac{1}{2} \sqrt{3} \ln \left(\frac{1+c^2 R^2}{x_4^2}\right)
    \end{aligned}
\end{equation}
where we denote $(c \cdot dx ) = -c_1 dx^1 + c_2 dx^2 + c_3 dx^4 $.

\subsection{Uni-vector deformations as coordinate transformations}
\label{sec:monocoord}

Given the analysis above one concludes, that despite the complexity of uni-vector deformations written in the form \eqref{glDdeformedsolution1}, these are nothing but coordinate transformations in the full space-time of the higher dimensional parent theory, given $\a^m$ is a Killing vector. To show this explicitly and in general we start with the representation of the transformations in terms of the blocks of the higher-dimensional metric $\tilde{G}_{MN}$:
\begin{equation}
\label{eq:block_transform1}
    \begin{aligned}
        \tilde{G}_{mn} & = G_{mn} \\
        \tilde{G}_{mz} & =G_{mz} + G_{mn}\a^n\\
        \tilde{G}_{zz} & = G_{zz} + 2 G_{z m} \a^m + G_{mn}\a^m \a^n
    \end{aligned}
\end{equation}
Let us first show that this is a coordinate transformation for constant $\a^m$, that means 
\begin{equation}
    L_\a G_{mn}(x) = \a^{k}\dt_k G_{mn}(x) = 0,
\end{equation}
and similarly for other components. In other words, fields do not depend on the KK coordinate $z$.

Under the transformation \eqref{eq:block_transform1} the interval transforms as
\begin{equation}
    \begin{aligned}
    ds^2  & = \tilde{G}_{MN}(x)dx^M dx^N\\
    & = G_{mn}(x)(dx^m + \a^m dz)(dx^n + \a^n dz) + 2 G_{mz}(x)(dx^m + \a^m dz) dz + G_{zz}(x)dzdz.
    \end{aligned}
\end{equation}
The examples considered previously and the form of the interval suggest the following coordinate transformation
\begin{equation}
    x'{}^m = x^m + \a^m z \quad \Longrightarrow \quad dx'{}^m = dx^m + \a^m dz,
\end{equation}
that gives for the interval
\begin{equation}
    ds^2  = G_{mn}(x)dx'{}^m dx'{}^n + 2 G_{mz}(x)dx'{}^m dz + G_{zz}(x)dzdz.
\end{equation}
Since the Killing vector is constant, and does not have $z$ component, and the metric does not depend on $z$ we have
\begin{equation}
    G_{mn}(x') = G_{mn}(x+ \a z) = G_{mn}(z) + \a^k\dt_k G_{mn}(x) + \dots = G_{mn}(x),
\end{equation}
and similarly for the other components. Hence, for the interval we derive
\begin{equation}
    \label{eq:wat}
    ds^2 = G_{mn}(x')dx'{}^m dx'{}^n + 2 G_{mz}(x')dx'{}^m dz + G_{zz}(x')dzdz.
\end{equation}
We obtained the same interval as prior to the deformation, meaning that the transformation can be undone by a coordinate transformation, provided $\a^m$ is a (constant) Killing vector. 

To see the same in a general case one has to go through more calculations, which however are straightforward. The corresponding coordinate transformation can be conveniently defined as
\begin{equation}
\label{eq:unicoord}
    x'{}^M = e^{ \x }x^M,
\end{equation}
where $\x = z \a^m \dt_m$ acts as a shift operator. To derive $dX'^M=(dx'{}^m,dz')$ we first notice
\begin{equation}
    dX'{}^M = \fr{\dt X'{}^M}{\dt X^N} dX^N,
\end{equation}
or in split components
\begin{equation}
    \begin{aligned}
        dx'{}^m & = \fr{\dt x'{}^m}{\dt x^n} dx^n + \fr{\dt x'{}^m}{\dt z}dz ,\\
        dz' & = \fr{\dt z'}{\dt x^n} dx^n + \fr{\dt z'}{\dt z}dz .
    \end{aligned}
\end{equation}
The transformation matrix can be rewritten in the following nice form (see Appendix \ref{app:coord} for more details)
\begin{equation}
    \fr{\dt X'{}^M}{\dt X^N} = \left( e^{(\x \mathbb{1} + a)}\mathbf{1}\right){}_N{}^M,
\end{equation}
where $\mathbf{1}$ denotes action on the function that simply equal to unity, $\mathbb{1}$ is the unity matrix and
\begin{equation}
    a_N{}^M = \dt_N (\a^M z).
\end{equation}
We intentionally follow the same notations as in \cite{Hohm:2012gk} so that the algorithm can be translated to double field theory coordinate transformations. 

Now, for components of $a_N{}^M$ we have
\begin{equation}
    a_N{}^M = 
        \begin{bmatrix}
            z \dt_n \a^m & \a^m \\
            0 & 0
        \end{bmatrix},
\end{equation}
that immediately implies
\begin{equation}
    \begin{aligned}
        e^{(\x \id+ a)}\mathbf{1}{}_z{}^z = 1, && e^{(\x \id+ a)}\mathbf{1}{}_m{}^z = 0,
    \end{aligned}
\end{equation}
from which we derive $dz' = dz$. The remaining components appear to be of the form
\begin{equation}
    \begin{aligned}
        e^{(\x \id + a)}\mathbf{1}{}_n{}^m & = \left(e^{\x \id+a}\right){}_n{}^me^{-\x}, \\
        e^{(\x \id + a)}\mathbf{1}{}_z{}^m &=e^{(\x \id + a)}\mathbf{1}{}_n{}^m \a^n  =\big(\a\, e^{\x \id+a}e^{-\x}\big)^m.
    \end{aligned}
\end{equation}
Let us now show that $\left(e^{\x \id + a}\right){}_n{}^m$ can be understood as a generator of translations, and hence of the Lie derivative, in the lower dimensional theory. For that we write 
\begin{equation}
    \left(e^{\x \id+a}\right){}_n{}^m = \sum_{p=0}^{\infty} \fr1{p!}
    (\x \id + a)_n{}^{M_1} \dots (\x \id + a)_{M_{p-2}}{}^{M_{p-1}}(\x \id + a)_{M_{p-1}}{}^{m}.
\end{equation}
Consider first the case when $M_{p-1} = z$ in the sum, then the previous factor is non-zero only if $M_{p-2}=z$ and so on until we end up with $n$, that is never $z$. Therefore, for any $p$ the sum can never include terms with the index $z$ at any place and we may understand $z$ in $\left(e^{\x \id+a}\right){}_n{}^m $ as simply a parameter. Altogether we have for the coordinate transformations
\begin{equation}
    \begin{aligned}
        dx'{}^m & = \left(e^{\x \id+a}\right){}_n{}^m  e^{-\x} \big(dx^n + \a^n(x) dz\big), \\
        dz' & = dz.
    \end{aligned}
\end{equation}

Consider now a term in the interval 
\begin{equation}
    \begin{aligned}
        G_{mz}(x)(dx^m + \a^m dz) & = G_{mz}(x)e^{\x}e^{-(\x \id+a)}{}_k{}^m e^{(\x \id+a)}{}_n{}^k e^{-\x} (dx^n + \x^n dz)\\
        & = e^{\x}e^{-(\x \id+a)}{}_k{}^m G_{mz}(x)dx'{}^k 
        = e^{\x}e^{-L_\x}G_{kz}(x)dx'{}^k \\
        & = e^{\x}G_{mz}(x) dx'{}^m = G_{mz}(x')dx'^m.
    \end{aligned}
\end{equation}
Here in the first line we simply inserted a Kronecker delta $\d_n{}^m$ in the form of an operator  and its inverse. In the second line we introduced $dx'{}^m$, and used the fact that $e^{\x}e^{-(\x \id+a)}$ is simply a function and can be moved to the very front. Next we recall the Killing vector property
\begin{equation}
   e^{-(\x+a)}A_{m}(x) = e^{-z L_\a}A_m(x) =  A_{m}(x),
\end{equation}
and finally use the definition of $x'{}^m$ as $e^\x$ acting on $x^m$. The same lines can be repeated for terms with $G_{mn}$ and $G_{zz}$ to show that our field redefinition is nothing but a coordinate transformation in the higher dimensional parent theory. 

It is worth to mention that no Yang--Baxter-like condition has appeared in the procedure both in the field redefinition and coordinate transformation approaches. As we will see in the next section this is directly related to the fact that the full higher dimensional space-time carries the structure of standard diffeomorphisms and the introduced coordinate transformations form a closed algebra without additional requirements, This will no longer be the case for bi-vector deformations.

\section{Bi-vector Yang-Baxter deformations}\label{sec:bi-vector}

The transformations parametrized by a single Killing vector field (which we call uni-vector deformations for homogeneity of notations) described above can be generalized to poly-vector deformations, i.e. those parametrized by a rank $p$ poly-vector $\Omega^{m_1\dots m_p}$. The simplest case $p=2$ gives bi-vector deformations that under certain conditions are, as we discuss below, precisely Yang--Baxter bi-vector deformations. In terms of the supergravity fields of the NS-NS sector providing a consistent string background (up to adding R-R fields and the dilaton) a bi-vector deformation is given by the following simple formula
\begin{equation}
    G + B = \big((g+b)^{-1} + \b\big)^{-1}.
\end{equation}
Here $g_{mn}$ and $b_{mn}$ are the initial metric the metric and the Kalb--Ramond 2-form field, $G_{mn}$ and $B_{mn}$ are the deformed fields, and $\b^{mn}=-\b^{nm}$ is the bi-vector. In order for such a transformation to map a solution into a solution one imposes the following conditions
\begin{equation}
    \label{eq:cybeuni}
    \begin{aligned}
        \b^{l[m}\dt_l \b^{nk]}&=0,
        \dt_m \b^{mk} &= 0.
    \end{aligned}
\end{equation}
In the bi-Killing ansatz $\b^{mn} = r^{A_1 A_2}k_{A_1}{}^mk_{A_2}{}^n$, where $r^{A_1A_2}= - r^{A_2 A_1} = $ const,  the first line above becomes classical Yang--Baxter equation and the second line become the so-called unimodularity condition 
\cite{Bakhmatov:2017joy,Bakhmatov:2018bvp,Bakhmatov:2022lin,Borsato:2016ose}
\begin{equation}
\label{10dcond}
    \begin{aligned}
         f_{B_1 B_2}\,^{[A_1} r^{A_2|B_1|} r^{A_3]B_2} &= 0, \\
         r^{A_1 A_2}f_{A_1 A_2}\,^{B}  & = 0.
    \end{aligned}
\end{equation}
Here $f_{A_1 A_2}{}^{A_3}$ denote structure constants of the algebra of Killing vectors $k_{A}{}^m$. 

\subsection{Bi-vector deformations in supergravity}

Let us now briefly review the standard approach to bi-vector deformations in 10-dimensional supergravity to compare with the uni-vector deformations above. Let us start with the action for the NS-NS sector of 10-dimensional supergravity $e_{m}{}^{a}$, $b_{m n}$ and $\phi$ 
\begin{equation}\label{DFT10sugraact}
S = \int d^{d} x \sqrt{|g|} e^{- 2 \phi} \left( R + 4 \dt_m \phi \dt^{m} \phi - \frac{1}{12} H_{m n k} H^{m n k} \right),
\end{equation}
where $H_{m n k} = 3 \dt_{[m} b_{n k]}$. This theory can be rewritten in an $\rmO(10,10)$ covariant framework called double field theory (DFT) and defined by the following action
\begin{equation}\label{DFTactionQ}
    S_{\text{DFT}} = \int d X e^{-2d} \mathbb{Q},
\end{equation}
where
\begin{equation}
    \mathbb{Q} = \mH^{\mA\mB} \mF_\mA \mF_\mB + \mF_{\mA\mB\mC}\mF_{\mD\mE\mF}\left(\fr14\mH^{\mA\mD}\h^{\mB\mE}\h^{\mC\mF} - \fr{1}{12}\mH^{\mA\mD}\mH^{\mB\mE}\mH^{\mC\mF}\right) - \mF_\mA \mF^\mA - \fr16 \mF_{\mA\mB\mC}\mF^{\mA\mB\mC}.
\end{equation}
The measure $dX$ denotes formal integration over 20 coordinates $X^\mM = (x^m,\tilde{x}_m)$, of which all fields will depend only on the so-called geometric coordinates $x^m$. The so-called dual coordinate $\tilde{x}_m$ will be crucial when defining bi-vector deformations as coordinate transformations in the full doubled space. The invariant $\rmO(10,10)$ tensor $\eta$ and flat generalized metric $\mH$ are given by
\begin{equation}\label{DFTflatetametric}
    \eta_{\mA\mB} = 
        \begin{pmatrix}
		       	0 & \delta_{a}{}^{b} \\
		    	\delta^{a}{}_{b} &  0
	    \end{pmatrix}, \quad
     \mH_{\mA \mB} = 
        \begin{pmatrix}
		       	h_{a b} & 0 \\
		    	0 &  h^{a b}
	    \end{pmatrix}.
\end{equation}
The flat metric $h_{ab}$ relates the vielbein and the metric as $g_{mn}= e_m{}^a e_n{}^b h_{ab}$. The fluxes are given in terms of the generalized vielbein $E_\mM{}^\mA$ and read
\begin{equation}\label{DFTfluxes}
    \mF_{\mA \mB \mC} = 3 E_{[\mC| \mN} \partial_{|\mA} E_{\mB]}{}^{\mN},\qquad \mF_{\mA} = 2 \partial_{\mA} d - \partial_{\mM}{E_{\mA}{}^{\mM}},
\end{equation}
where $\dt_{\mA} = E_{\mA}{}^{\mM} \dt_{\mM}$ and $\dt_\mM$ denotes derivative w.r.t. $\XX^\mM$, indices are raised and lowered by the invariant tensors $\h_{\mM\mN}$ and $\h_{\mA\mB}$ (with curved and flat indices respectively). The last ingredient needed for consistency is the so-called section condition
\begin{equation}\label{2condDFT2}
    \eta^{\mM \mN} \partial_{\mM} f \, \partial_{\mN} \, g = 0, \quad \eta^{\mM \mN} \partial_{\mM} \, \partial_{\mN} \, f = 0,
\end{equation}
that must hold for any fields $f$ and $g$ of the theory. In order to reproduce the action and equations of motion for the NS-NS sector of 10-dimensional supergravity we need to parametrize DFT fields as follows
\begin{equation}\label{DFTdinsertion}
        d = \phi - \frac12 \log\,\det\,e^{a}{}_{m},
\end{equation}
\begin{equation}\label{DFTeinsertion}
            E^{A}{}_M = 
            \begin{pmatrix}
            e^{a}{}_m && - e_{a}{}^{k} b_{k m} \\ \\
            0 && e_{a}{}^{m}
            \end{pmatrix}, \quad
            E_{A}{}^M = 
            \begin{pmatrix}
            e_{a}{}^m && 0 \\ \\
            - e_{a}{}^{k} b_{k m} && e^{a}{}_m
            \end{pmatrix},
\end{equation}
where $e = \det\, e^{a}{}_{k}$, $g_{m n} = e^{a}{}_{m} e^{b}{}_{n} g_{a b}$. Moreover, we must fix the solution of (\ref{2condDFT2})
\begin{equation}\label{sectionDFTsolution}
            \partial_{\mM} = (\partial_{m} , \, \tilde{\partial}^{m}), \quad \tilde{\partial}^{m} \equiv 0.
\end{equation}
In the present context the coordinates $\tilde{x}_{m}$ will be analogous to the coordinate $x^{D+1}$ used for the KK reduction previously.

Bi-vector deformations are $\rmO(10,10)$ transformation that act on the generalized vielbein as follows
\begin{equation}\label{rotation}
    E'^{\mA}{}_{\mM} = O_{\mM}{}^{\mN} E^{\mA}{}_{\mN} ,
\end{equation}
with the matrix given by
\begin{equation}
            O_\mM{}^\mN = \begin{pmatrix}
            \d_m{}^n && 0 \\
            \\
            - \beta^{m n} && \d^m{}_n
            \end{pmatrix}.
\end{equation}
The deformed generalized vielbein is no longer upper triangular:
\begin{equation}
            E'^{\mA}{}_\mM = 
            \begin{pmatrix}
            e^{a}{}_m && - e_{a}{}^{k} b_{k m} \\ \\
            - \beta^{m n} e_{a}{}^{n} && \beta^{m n} e_{a}{}^{k} b_{k n} + e_{a}{}^{m}
            \end{pmatrix},
\end{equation}
and in general requires an $\rmO(10)\times \rmO(10)$ transformation to reproduce the standard supergravity fields. Instead, one composes the generalised metric $\mH_{\mM\mN}$ and reads deformed $G_{mn}$ and $B_{mn}$ directly. Invariance of the action in the bi-Killing ansatz, $\beta^{m n} = r^{AB} k_{A}{}^{m} k_{B}{}^{n}$, is achieved by requiring
\begin{equation}
\begin{aligned}
        \mL_{k_{A}} \, \phi = 0, \quad \mL_{k_{A}} \, g_{mn} = 0, \quad \mL_{k_{A}} \, b_{m n} = 0,
\end{aligned}
\end{equation}
and the algebraic constraints \eqref{10dcond}. Let us again mention the issue, that vanishing Lie derivative on the metric does not necessarily vanishes on the vielbein. Careful consideration shows that fluxes do transform under a general bi-vector deformation, however, the action is still invariant and all relevant statements hold.

\subsection{Bi-vector deformations as diffeomorphisms in extended (doubled) space}\label{sec:bi-vector-coord}

We would like to show that bi-vector Yang--Baxter unimodular deformations are nothing but a coordinate transformation in the so-called doubled space. Moreover the conditions \eqref{eq:cybeuni} appear as a consequence of the requirement that the transformation belongs to the orthogonal group. For that we consider a space of dimension $20$ parametrized by coordinates $X^M= (x^m,\tx_m)$, where the first ten $x^m$ are referred to as geometric and are used to measure physics distances in the standard space-time, and the latter ten $\tx_m$ are usually called dual and are related to winding modes of closed strings. Such a doubled space naturally appears in the closed string field theory framework \cite{Siegel:1993xq,Siegel:1993th} and is a consequence of describing left and right moving modes of the closed string as independent. In the framework of double field theory dynamics of supergravity is described in generalized geometrical terms similar to those of general relativity \cite{Hohm:2010pp}. Without digging more into these structure let us mention that all fields in our setup will depend on $x^m$ only.

Consider now the following coordinate transformations in the doubled space-time 
\begin{equation}
    X'{}^M = e^{\x} X^M,
\end{equation}
where 
\begin{equation}
    \x = \b^{m,n}\tx_m \dt_n.
\end{equation}
In contrast to the uni-vector case one immediately faces the issue that for such transformations to form a closed algebra of diffeomorphisms in the ordinary space-time parametrized by $x^m$ additional conditions are required. To see that, first note that as in the previous case the dual coordinate $\tx_m$ plays the role of the transformation parameter and hence it is convenient to denote $\x_{\tx} =\b^{m,n}\tx_m \dt_n$. For the algebra to close we need
\begin{equation}
    [\d_{\x_{\tx}}, \d_{\x_{\ty}}] = \d_{[\x_{\tx},\x_{\ty}]} = \d_{\x_{\tilde z}},
\end{equation}
where the parameter $\tz_m$ is composed of $\tx_m$, $\ty_m$ and  $\b^{m,n}$, and the bracket is the ordinary Lie derivative. Hence, we write
\begin{equation}
    \begin{aligned}
        [\d_{\x_x}, \d_{\x_y}] & = \b^{m,k}\tx_m \dt_k(\b^{nl}\ty_n \dt_l) - (\tx \leftrightarrow \ty) \\
        & = 2 \big( \b^{m,k}\dt_k \b^{n,l} \big)\tx_{[m} \ty_{n]} \dt_l + 2\b^{m,k}\b^{n,l}\tx_{[m} \ty_{n]}\dt_k \dt_l.
    \end{aligned}
\end{equation}
The last term is apparently zero, while the first must be rewritten in the form $\b^{k,l}\tz_k \dt_l$. Although in general $\b^{m,n}$ is not required to have a particular symmetry under permutation of indices, for simplicity we assume
\begin{equation}
    \b^{m,n} = -\b^{n,m}
\end{equation}
and in what follows use the notation $\b^{mn}$ (without separation of the indices). Now, requiring the condition
\begin{equation}
    \label{eq:Rflux}
    \b^{l[m}\dt_l \b^{nk]}=0,
\end{equation}
we obtain
\begin{equation}
    [\d_{\x_x}, \d_{\x_y}] =  \b^{kl}(\dt_k \b^{mn}) \tx_m \ty_n \dt_l,
\end{equation}
that is of the desired form.

An important difference with the previous case of uni-vector deformations is that the ansatz for $\x^m$  taken as above does not quite work. Technically the reason is that at some point the derivative $L_\x$ appears that is say for a co-vector $V_m$ gives
\begin{equation}
    \begin{aligned}
        L_\x V_m & = \x^n\dt_n V_m + V_n\dt_m \x^n \\
        &= r^{AB}\left(\tx_n k_A{}^n L_{k_B}V_m + \tx_k V_{n}\dt_m k_A{}^k  k_B{}^n\right).
    \end{aligned}
\end{equation}
The first term in parentheses vanishes since $k_A^m$ is a Killing vector, while the second is in general non-zero. Fundamentally the reason might be related to the fact that the dual coordinates $\tx_m$ carry the same index as a normal geometric tensor, while do not transform on themselves. This leads to a lack of factors of $\dt x'^m/\dt x^n$ and results in incorrect transformations. Nevertheless, in Section \ref{sec:almost} by considering a special case of almost-abelian deformations classified in \cite{Borsato:2018idb}, and  using large coordinate transformation of the doubled space found in \cite{Hohm:2012mf} we show that the procedure works as in the uni-vector case. Before turning to this special class of transformations let us develop a more general picture that motivates using of such a special ansatz.

To start with we notice that tensors of double field theory under a coordinate transformation $X'{}^M = X'{}^M(X)$ transform by the $\rmO(10,10)$ matrix \cite{Hohm:2012mf}
\begin{equation}
\label{eq:largecoordF}
    \mF_\mM{}^\mN = \fr12 \left( \fr{\dt X^\mP}{\dt X'{}^\mM}\fr{\dt X'_\mP}{\dt X_\mN} 
    + \fr{\dt X'_\mM}{\dt X^\mP}\fr{\dt X^\mN}{\dt X'_\mP} \right).
\end{equation}
For the generalized metric we then have
\begin{equation}
\label{eq:largecoord}
    \mH'(X')_{\mM\mN} = \mF_{\mM}{}^{\mK}\mF_{\mN}{}^{\mL}\mH_{\mK\mL}(X).
\end{equation}
Note also that the construction $ds^2 = \mH_{\mM\mN}dX^\mM dX^\mN$ is not invariant under general coordinate transformation in the doubled space and hence cannot be used for our purposes.

Consider now a general coordinate transformation in the full doubled space at the linear level to arrive at requirements for the transformation to reproduce the rules of a bi-vector Yang--Baxter deformation. The latter read
\begin{equation}
\label{eq:bivecdef}
    \begin{aligned}
        \mH_{mn}(x)' & = \mH_{mn}(x), \\
        \mH_m{}^n(x)' & = \mH_m{}^n(x) + \b^{nk}\mH_{mk}, \\
        \mH^{mn}(x)' & = \mH^{ mn}(x) + 2\b^{(m|k|}\mH_{k}{}^{n)} 
        + \b^{mk}\b^{nl}\mH_{kl}.
    \end{aligned}
\end{equation}
Important is that the coordinate point on the LHS and RHS is the same, i.e. in this form the deformation is a transformation of fields. To write it as a doubled coordinate transformation consider
\begin{equation}
    X'{}^\mM = X'{}^\mM + \d^\mM(X),
\end{equation} 
where $\d^M = (\d^m, \td_m)$ is a general coordinate shift. At the linear level this is simply a shift by generalized Lie derivative along $\d^\mM$
\begin{equation}
    \begin{aligned}
    \mH_{\mM\mN}'(X) = \mH_{\mM\mN} - \left(\d^\mK \dt_\mK \mH_{\mM\mN}  +2 \left( \dt_{(\mM}\d^\mK - \dt^\mK \d_{(\mM}\right) \mH_{\mN)\mK}\right).
    \end{aligned}
\end{equation}
Comparing this to \eqref{eq:bivecdef} we derive the following conditions for components of the vector $\d^M$
\begin{equation}
    \begin{aligned}
        \tdt^m \d^n - \tdt^n \d^m & = 2 r^{AB}k_A{}^mk_B{}^n,\\
        \dt_m \td_n - \dt_n \td_m & = 0,\\
        \d^k \dt_k \mH_{mn} + 2(\dt_{(m} \d^k - \tdt^k \td_{(m})\mH_{n)k} & = \#^A L_{k_A} \mH_{mn} = 0.    
    \end{aligned}
\end{equation}
The last line is the same for the components $\mH^{mn}$ and $\mH_m{}^n$ and the hash sign denotes any expression. From the first line we learn that $\d^m$ does not necessarily need to be proportional to the antisymmetrized contraction of Killing vectors --- the antisymmetrization comes for free from the structure of the transformation. This is actually not a surprise since the corresponding generator is parametrized by an antisymmetric pair of indices, that gives
\begin{equation}
    \d^m = \bar{r}^{A,B}k_A{}^m k_B{}^n + \tdt^m \d,
\end{equation}
where $\d$ is some function of both $x^m$ and $\tx_m$ and $\bar{r}^{A,B}$ has no particular symmetry in indices. Similarly the second line implies $    \td_m = \dt_m \td$. Now we turn to the last line, where we face the same problem as before, that originates from the terms where derivatives $\dt_m$ hit on the first Killing vector $k_A^m$ in the product:
\begin{equation}
    \tdt^k \d \dt_k \mH_{mn} + 2 \bar{r}^{A,B} k_N{}^k \tx_l \dt_{(m} k_A{}^l \, \mH_{n)k} + 2\mH_{k(m}\dt_{n)} \tdt^k \D\,  ,
\end{equation}
where $\D = \d - \td$. Since coordinate transformations here can be constructed only of the Killing vectors, the simplest expression for $\d$ that we can write at the level linear in $\bar{r}^{A,B}$ is
\begin{equation}
    \d = \a\, \bar{r}^{A,B} \tx_k \tx_l k_A{}^k k_B{}^l,
\end{equation}
where $\a$ is a numerical coefficient. Certainly one may add terms with derivatives of $k$, which however will lead to terms with two derivatives to cancel which one will have to introduce terms with two derivatives  and so on. With this simplest choice of $\d$ we get
\begin{equation}
    4 \bar{r}^{(A,B)} \tx_l \left(k_B{}^k\dt_m k_A{}^l - \a k_A{}^l \dt_m k_B{}^k \right) \mH_{nk} + 2 \mH_{nk}\dt_m \tdt^k \D - 2 \bar{r}^{A,B} k_A{}^k \tx_l \dt_m k_B{}^l \, \mH_{nk},
\end{equation}
where symmetrization in $m$, $n$ is undermined. We see that setting $\a= -1$ allows to rewrite terms in the parentheses as a full derivative $\dt_m \tdt^k$ and hide them into $\td$. The remaining term is apparently not of this form in general and no choice of $\a$ allows it to fit in. However, the form of the last term together with the issues recovered at earlier steps suggest a solution to this problem that nicely describes almost all non-abelian Yang--Baxter deformations found in \cite{Borsato:2016ose}. Indeed, suppose the full set of Killing vectors is decomposed into $k_a^m$, that commute and can be made constant, and $k_\a^m$, that depends on $x^m$ linearly. Let the constant matrix has only components $\bar{r}^{a,b}$ and $\bar{r}^{a,\a}$. Then the last term does not depend on $x^m$ and, as we show below, can be undone by a TsT transformation. What we describe here fit into a particular subset of almost-abelian deformation, to whose analysis we turn immediately.

\subsection{Almost abelian deformations are coordinate transformations}
\label{sec:almost}

Let us consider now deformations with bi-vectors of the following form
\begin{equation}
    \b = p_1 \wedge p_2 + q \wedge j,
\end{equation}
where the only non-vanishing commutators are
\begin{equation}
    [j,p_i] = \e_i q.
\end{equation}
Deformations in this form fit a large subset of the non-abelian deformations classified in \cite{Borsato:2016ose}. Since $p_1$, $p_2$ and $q$ all commute, they can be chosen independent of coordinates, while $j$ depends on $x^m$ linearly as a momentum generator. This allows us to derive the following relations that will be actively used in what follows
\begin{equation}
    p_{i}{}^{m} \dt_{m} j^{n} = - \e_{i} q^{n}, \quad q^{k} \dt_{k} j^{n} = 0, \quad \dt_{k} q^{n} = 0, \quad \dt_{k} p_{i}{}^{n} = 0.
\end{equation}
It also proves convenient to introduce the following notations
\begin{equation}
    \xi = (\underbrace{\tilde{x}_l p_1{}^l p_2{}^{m}}_{\xi_1{}^m} + \underbrace{\tilde{x}_l q^l j^{m}}_{\xi_2{}^m}) \dt_{m} = (\xi_1{}^m + \xi_2{}^m) \dt_{m} = \xi_1 + \xi_2.
\end{equation}
In $\rmO(10,10)$ covariant expressions we will use the same notations, but for $\xi_i = \xi_i{}^\mM \dt_{\mM}$ assuming $\xi_{i m} = 0$. We denote the result of commutation of $\x_1$ and $\x_2$ by $\l$:
\begin{equation}
    [\xi_1,\xi_2] = - \e_2 \underbrace{\tilde{x}_l p_1{}^l}_{\a} \underbrace{\tilde{x}_k q^k}_{z} q^{n} \dt_{n} = - \e_2 \a z q^{n} \dt_{n} = - 2 \e_2 \l
\end{equation}
Now with the help of Zassenhaus formula we identify coordinate transformation
\begin{equation}
    X'^\mM = e^{\xi} X^{\mM} = e^{\xi_1 + \xi_2} X^{\mM} = e^{\xi_1} e^{\xi_2} e^{\e_2\l} X^{\mM} = e^{\e_2\l} e^{\xi_1} e^{\xi_2} X^{\mM}.
\end{equation}
Let us summarize our definitions as follows
\begin{equation}
\begin{aligned}
    \xi_2{}^m & = \tilde{x}_l q^l j^{m}, &&\dt_{N} \xi_2^M = a_{N}{}^{M},  &&a^{nm} = q^{n} j^m,  &&a_{n}{}^{m} = z \dt_{n} j^m \\
    \xi_1{}^m & = \tilde{x}_l p_1{}^l p_2{}^{m},  &&\dt_{N} \xi_1^M = b_{N}{}^{M}, &&b^{nm} = p_1{}^{n} p_2{}^m, &&b_{n}{}^{m} = 0,\\
    \l^m & = \frac12 \a z q^{m}, &&\dt_{N} \l^M = c_{N}{}^{M}, \quad &&c^{nm} = \frac12 (z p_1{}^n + \a q^n) q^m, &&c_{n}{}^{m} = 0.
\end{aligned}
\end{equation}
Note, that all components except $a^{nm}$ do not depend on $x^{m}$, so they commute with derivatives $\dt_m$. Derivative of $X'{}^\mM$ w.r.t. $X^\mM$ that defines transformations of tensors can be written in terms of the matrices introduced above:
\begin{equation}
\begin{aligned}
     \frac{\dt X'^\mM}{\dt X^{\mN}} &= \dt_{\mN}(e^{\e_2\l} e^{\xi_1} e^{\xi_2} X^{\mM}) = \sum_{r,s,t=0}^{\infty} \frac{1}{r!s!t!} \e_2{}^r [(\l + c)^{r} (\xi_1 + b)^{s} (\xi_2 + a)^{t} \, \mathbb{1}]_{\mN}{}^{\mM}  \\
    & = \sum_{r,s,t=0}^{\infty}  \frac{1}{r!s!t!} \e_2{}^r [(\l \d_{\mN}{}^{\mM_2} + c_{\mN}{}^{\mM_2}) \ldots (\l \d_{\mM_r}{}^{\mN_1} + c_{\mM_r}{}^{\mN_1})(\xi_1 \d_{\mN_1}{}^{\mN_2} + b_{\mN_1}{}^{\mN_2})\times\ldots\\
    &\times  \ldots (\xi_1 \d_{\mN_s}{}^{\mK_1} + b_{\mN_1}{}^{\mK_1})(\xi_2 \d_{\mK_1}{}^{\mK_2} + a_{\mK_1}{}^{\mK_2}) \ldots (\xi_2 \d_{\mK_t}{}^{\mM} + a_{\mK_t}{}^{\mQ})  \, \mathbb{1}_{\mQ}{}^{\mM}].
\end{aligned}
\end{equation}
One immediately has
\begin{equation}
     \frac{\dt \tilde{x}'_m}{\dt \tilde{x}_n} = \d_{m}{}^{n}, \quad \frac{\dt \tilde{x}'_m}{\dt x^n} = 0.
\end{equation}
Before proceeding with the remaining components we notice the following statement 
\begin{equation}
     (e^{\e_2 \l} e^{\xi_1} e^{\xi_2 + a})_{n}{}^{m} = (e^{\e_2 \l} e^{\xi_1} e^{\xi_2 + a} \mathbf{1})_{n}{}^{m} e^{\e_2\l} e^{\xi_1} e^{\xi_2},
\end{equation}
that is a simple consequence of \eqref{eq:op2fun}, and we drop $\mathbb{1}$ from the exponent for clarity of notations. Using this we obtain
\begin{equation}\label{xoldxnewder}
     \frac{\dt x'^m}{\dt x^{n}} = (e^{\e_2\l} e^{\xi_1} e^{\xi_2 + a} \mathbf{1})_{n}{}^{m} = (e^{\e_2\l} e^{\xi_1} e^{\xi_2 + a})_{n}{}^{m} e^{-\xi_2} e^{-\xi_1} e^{-\e_2\l} = (e^{\xi + a})_{n}{}^{m} e^{-\xi},
\end{equation}
in the last one we used $\l a_{m}{}^{n} = 0$ and $\xi_{1} a_{m}{}^{n} = 0$ and Zassenhaus formula. The last and the most non-trivial relation is
\begin{multline}
     \frac{\dt x'^m}{\dt \tilde{x}_{n}} = [e^{\e_2\l} e^{\xi_1} (b^{nq} + \e_2c^{nq} + a^{nq}) (e^{\xi_2 + a})_{q}{}^{m} \mathbf{1}] =\\
     = (b^{nq} + \e_2c^{nq} + a^{nq} + \e_2(\l a^{nq}) + (\xi_1 a^{nq})) (e^{\e_2\l} e^{\xi_1} e^{\xi_2 + a} e^{-\xi_2} e^{-\xi_1} e^{-\e_2 \l})_{q}{}^{m} =\\
     = (e^{\e_2\l} e^{\xi_1} e^{\xi_2 + a} e^{-\xi_2} e^{-\xi_1} e^{-\e_2\l})_{q}{}^{m} (b^{nq} + \e_2c^{nq} + a^{nq} + \e_2(\l a^{nq}) + (\xi_1 a^{nq})).
\end{multline}
The following simplifications are in place
\begin{equation}
    \begin{aligned}
        (\l a^{mn}) & = \fr12 \a z q^k \dt_k(q^m j^n) = \fr12 \a z q^m [q,j]^n =0, \\
        (\x_1 a^{mn}) & = \a p_2{}^k \dt_k (q^m j^n) = \a q^m [p_2,j]^n = - \a \e_2 q^m q^n, 
    \end{aligned}
\end{equation}
that give the final result summarized as:
\begin{equation}
\begin{aligned}
    d x'^m & = (e^{\e_2\l} e^{\xi_1} e^{\xi_2 + a} e^{-\xi_2} e^{-\xi_1} e^{-\e_2\l})_{n}{}^{m} (dx^n + (b^{kn}  + a^{kn}+ \fr12\e_2 (z p_{1}{}^k q^n - \a q^n q^k)) d \tilde{x}_k),\\
    & = \fr{\dt x'{}^m}{\dt x^n}dx^n + \fr{\dt x'{}^m}{\dt x^n}\left(b^{kn}  + a^{kn}+ \fr12\e_2 (z p_1{}^k q^n - \a q^n q^k)\right) d \tilde{x}_k \\
    & = \fr{\dt x'{}^m}{\dt x^n}dx^n + \fr{\dt x'{}^m}{\dt x^n} A^{k n} d \tilde{x}_k \\
    d \tilde{x}'_m & = d \tilde{x}_m.
\end{aligned}
\end{equation}
Here we introduce a shorthand notation for 
\begin{equation}
    A^{mn} = b^{mn} + a^{mn} + \fr12\e_2\big(z p_1{}^m q^n - \a q^m q^n\big),
\end{equation}
that will be used later.

Components of the inverse matrix  $\dt X^\mM/\dt X'{}^\mN$ can be found easily from its definition that is in components
\begin{equation}
    \begin{aligned}
        \fr{\dt x'{}^m}{\dt x^k}\fr{\dt x^k}{\dt x'{}^n} &= \d_n{}^m, && \fr{\dt \tilde{x}_m}{\dt x'{}^n} = 0 \\
        \fr{\dt \tilde{x}_m}{\dt \tilde{x}'_n} &= \d_m{}^n, &&   
        \fr{\dt x^n}{\dt \tilde{x}'_k} = - A^{kn}.
    \end{aligned}
\end{equation}
Note the order of indices in the last expression.

For transformation of generalized tensors on the doubled space we need the matrix 
\begin{equation}
    \mF_\mM{}^\mN = \fr12 \left( \fr{\dt X^\mP}{\dt X'{}^\mM}\fr{\dt X'_\mP}{\dt X_\mN} 
    + \fr{\dt X'_\mM}{\dt X^\mP}\fr{\dt X^\mN}{\dt X'_\mP} \right).
\end{equation}
Using the result for $\dt X^\mM/\dt X'{}^\mN$ and  $\dt X'^\mM/\dt X{}^\mN$ a straightforward calculation gives
\begin{equation}
    \begin{aligned}
        \mF_m{}^n & = \fr{\dt x^n}{\dt x'{}^m},\quad \mF^m{}_n = \fr{\dt x'{}^m}{\dt x^n}, \quad \mF_{mn} = 0\\
        \mF^{mn} &= \fr{1}{2}\left(- A^{mn}  + A^{p k} \fr{\dt x'{}^m}{\dt x^k}\fr{\dt x{}^n}{\dt x'{}^p} + A^{nk}\fr{\dt x'{}^m}{\dt x^k} - 
         A^{kn}\fr{\dt x'{}^m}{\dt x^k}\right).
    \end{aligned}
\end{equation}

The idea of our derivation is to reproduce the bi-vector deformation rules \eqref{eq:bivecdef} from coordinate transformations \eqref{eq:largecoord}. For that we have to shift the LHS of \eqref{eq:largecoord} back to $X$, that is done simply by action of $e^{-\x}$. Hence, overall we want to calculate
\begin{equation}
    \mH'_{MN}(X) = e^{-\x}\mF_{M}{}^{K}\mF_{N}{}^{L}\mH_{KL}(X).
\end{equation}
For $\mH_{mn}$ we immediately arrive at the desired result (see also section 4 of \cite{Hohm:2012gk})
\begin{equation}
    \begin{aligned}
        \mH'_{mn}(x)=e^{-\x}\mH'_{mn}(x') =  e^{-\x}\fr{\dt x^k}{\dt x'{}^m} \fr{\dt x^l}{\dt x'{}^n}\mH_{kl}(x)  = e^{-L_\x} \mH_{mn}(x) = \mH_{mn}(x),
    \end{aligned}
\end{equation}
where $L_\x \mH_{mn} = 0$ has been used at the last step.   
For $\mH_m{}^n$ one has to work more, and we start with writing
\begin{equation}
\label{eq:H12}
    \begin{aligned}
    \mH'_m{}^n(x) & = e^{-\x}\Big(\fr{\dt x^k}{\dt x'{}^m}\fr{\dt x'{}^n}{\dt x^l}\mH_k{}^l(x) + \fr{\dt x^k}{\dt x'{}^m}\mF^{nl} \mH_{kl}\Big) \\
    & = e^{-L_\x} \mH_{m}{}^n(x) + e^{-\x}\fr{\dt x^k}{\dt x'{}^m}\mF^{nl}(x) \mH_{kl}(x)
    \end{aligned}
\end{equation}
Let us first take the second term in \eqref{eq:H12} and add a couple of Kronecker delta's. This move allows to rewrite the expression as follows
\begin{equation}
\label{eq:FH12}
    \begin{aligned}
        e^{-\x}\fr{\dt x^k}{\dt x'{}^m}\mF^{nl}(x) \mH_{kl}(x)&=e^{-\x}\Bigg(\fr{\dt x^k}{\dt x'{}^m}\mF^{nl}(x) 
        \fr{\dt x'{}^r}{\dt x^k}\fr{\dt x'{}^s}{\dt x^l}
        \fr{\dt x^p}{\dt x'{}^r}\fr{\dt x^q}{\dt x'{}^s}
        \mH_{pq}(x)\Bigg) \\
        &=\Bigg(e^{-\x}\mF^{nl}(x)\fr{\dt x'{}^p}{\dt x^l}\Bigg) e^{-L_\x}\mH_{mp}(x) =  \Bigg(e^{-\x}\mF^{nl}(x)\fr{\dt x'{}^p}{\dt x^l}\Bigg)\mH_{mp}(x).
    \end{aligned}
\end{equation}
Here in the second line the shift $e^{\x}$ is acting only inside the parentheses and we used the apparent property $e^{\x}f(x)g(x) = \big(e^\x f(x)\big)\big(e^\x g(x)\big)$ valid for any two functions $f(x)$ and $g(x)$.

To arrive at the bi-vector deformation formula one has to show that the expression in parentheses in \eqref{eq:FH12} must be simply $\beta^{np}$. To show that we expand this expression as
\begin{equation}
    e^{-\x}\mF^{nl}(x)\fr{\dt x'{}^p}{\dt x^l} = \fr12 e^{-\x} \Big( -A^{nl} \fr{\dt x'^p}{\dt x^l} + A^{pl} \fr{\dt x'^n}{\dt x^l}   + 2 A^{[rs]} \fr{\dt x'^p}{\dt x^r} \fr{\dt x'^n}{\dt x^s}\Big)
\end{equation}
and zoom into the last term, that gives
\begin{equation}
    \begin{aligned}
    &e^{-\x} A^{[rs]} \fr{\dt x'^p}{\dt x^r} \fr{\dt x'^n}{\dt x^s} = e^{-L_\x} A^{[pn]} = e^{-L_\x} \big(b^{[pn]}  + a^{[pn]}+ \fr12\e_2 z p_1{}^{[p} q^{n]}\big)\\
    &=A^{[pn]} - z \e_1 q^{[p}p_2{}^{n]} - z \e_2 p_1{}^{[p}q^{n]}.
    \end{aligned}
\end{equation}
To get the second line we first expand the exponent of the Lie derivative and calculate the powers explicitly. Using
\begin{equation}
    \begin{aligned}
        L_\x q^{m} & = 0\\
        L_\x{} p_{i}{}^{m}& = \e_{i} \tilde{x}_{l} q^{l} q^{m} = \e_{i} z q^{m},\\
        L_\x j^{m} & = - \e_{2} \tilde{x}_{l} p^{l}_{1} q^{m} = - \e_{2} \a q^{m},
    \end{aligned}
\end{equation}
we arrive at the following expressions
\begin{equation}
    \begin{aligned}
        L_\x b^{[mn]} & = L_{\x}(p_1{}^{[m} p_2{}^{n]}) = L_{\x} p_1{}^{[m} p_2{}^{n]} + p_1{}^{[m}L_{\x}  p_2{}^{n]} = z \e_1 q^{[m} p_2{}^{n]} + z \e_2 p_1{}^{[m}q^{n]}\\
        L_\x{}^2 b^{[mn]}& \sim q^{[m}q^{n]} = 0,\\
        L_\x a^{[mn]} & = L_{\x}(q^{[m} j^{n]}) = q^{[m} L_{\x}j^{n]} = - \e_{2} \a q^{[m} q^{n]} = 0.
    \end{aligned}
\end{equation}

To process the remaining two terms in \eqref{eq:FH12} we first insert a delta Kronecker and then reorganize the factors as follows
\begin{equation}
     \fr12 e^{-\x} \Big(-A^{nl} \fr{\dt x'^p}{\dt x^l} + A^{pl} \fr{\dt x'^n}{\dt x^l} \Big) = e^{-\x} \fr{\dt x'^{[p}}{\dt x^r}\fr{\dt x'{}^{n]}}{\dt x^l} \fr{\dt x^r}{\dt x'^s}A^{sl}.
\end{equation}
We see that the first two factors together with $e^\x$ compose exponent of the Lie derivative of $\fr{\dt x^r}{\dt x'^s}A^{sl}$ as before. To deal with the extra factor $\dt x^r/\dt x'^s$ we rewrite it as
\begin{equation}
    \fr{\dt x^r}{\dt x'^s}A^{sl} = \big(e^\x e^{-(\x +a)}\big){}_s{}^r A^{s l}
\end{equation}
Let us calculate this expression expanding the action of $\x$ and $a$ in the exponents on each of the terms in $A^{sl}$. For convenience let us recall explicit form of $A^{mn}$
\begin{equation}
    A^{mn} = b^{mn} + a^{mn} + \fr12\e_2\big(z p_1{}^m q^n - \a q^m q^n\big) = q^{m} j^{n} + p_{1}^{m} p_{2}^{n} + \fr12\e_2\big(z p_1{}^m q^n - \a q^m q^n\big).
\end{equation}
We start with the terms $q^m q^n$ and  $q^mj^n$. First we notice
\begin{equation}
    (\x \d_s{}^r+a_s{}^r)q^s = \x q^r + z q^s \dt_s j^r = q^r \x + z[q,j]^r  = q^r \x,
\end{equation}
that allows to rewrite the exponents as follows
\begin{equation}
    \begin{aligned}
    e^\x \big(e^{-(\x+a)}\big)_s{}^r q^s t^l &= e^\x \sum_{n=0}^{\infty}\fr{1}{n!}(-1)^n
    (\x+a)_{s_1}{}^r\dots (\x+a)_s{}^{s_n} q^s t^l\\
    &=e^\x q^r \sum_{n=0}^{\infty} \fr{1}{n!}(-1)^n \x^n t^l = q^re^{\x}e^{-\x}t^l = q^r t^l,
    \end{aligned}
\end{equation}
where $t^m$ can be $q^m$ or $j^m$. 

The remaining terms in $A^{sl}$ are $p_1{}^m p_2{}^n$ and $p_1{}^m q^n$. For these we first write
\begin{equation}
    (\x \d_s{}^r+a_s{}^r)p_1{}^s = \x p_1{}^r + z p_1{}^s \dt_s j^r = p_1{}^r \x + z[p_1,j]^r  = p_1{}^r \x - \e_1 z q^r.
\end{equation}
This gives 
\begin{equation}
    (\x + a)^n p_1 = p_1 \x^n - n \e_1 z q \x^{n-1}
\end{equation}
Dropping indices for clarity of notations we write
\begin{equation}
\begin{aligned}
    e^\x \sum_{n=0}^{\infty} \fr{(-1)^n}{n!}(\x+a)^n p_1 & = e^\x \Bigg(\sum_{n=0}^{\infty} \fr{(-1)^n}{n!}p_1 \x^n - \sum_{n=1}^{\infty} \fr{(-1)^n}{(n-1)!}\e_1 z q \x^{n-1}\Bigg) \\
    &=e^\x\big(e^{-\x}p_1 + z\e_1 e^{-\x} q\big) = p_1 +  z \e_1 q.
\end{aligned}
\end{equation}
Collecting everything together we obtain
\begin{equation}
    \fr{\dt x^r}{\dt x'^s}A^{sl} = A^{rl} + z \e_1 q^r p_2{}^l + \fr12 \e_1 \e_2 z^2 q^r q^l.
\end{equation}

Finally for the desired term in the transformation of $\mH_m{}^n$ we obtain
\begin{equation}
\begin{aligned}
    e^{-\x}\mF^{nl}(x)\fr{\dt x'{}^p}{\dt x^l} & = \fr12 e^{-\x} \Big( -A^{nl} \fr{\dt x'^p}{\dt x^l} + A^{pl} \fr{\dt x'^n}{\dt x^l}   + 2 A^{[rs]} \fr{\dt x'^p}{\dt x^r} \fr{\dt x'^n}{\dt x^s}\Big) \\
    & = e^{-\x} \Bigg(\fr{\dt x'^{[p}}{\dt x^r}\fr{\dt x'{}^{n]}}{\dt x^l} \fr{\dt x^r}{\dt x'^s}A^{sl} + A^{[rs]} \fr{\dt x'^p}{\dt x^r} \fr{\dt x'^n}{\dt x^s}\Bigg) \\
    & = e^{-\x} \fr{\dt x'^p}{\dt x^r} \fr{\dt x'^n}{\dt x^s}\Big(  2A^{[rs]}  + z \e_1 q^{[r}p_1{}^{s]}\Big) = e^{-L_\x}\Big( 2A^{[pn]}  + z \e_1 q^{[p}p_2{}^{n]}\Big)\\
    &=2 A^{[pn]} - 2 z \e_1 q^{[p}p_2{}^{n]} - 2 z \e_2 p_1{}^{[p}q^{n]} + z \e_1 q^{[p} p_2{}^{n]} - z^2 \e_1 \e_2 q^{[p} q^{n]} \\
    &= 2 (b^{[pn]} + a^{[pn]}) + z \e_2 p_1{}^{[p} q^{n]} - \a q^{[m} q^{n]}  - 2 z \e_1 q^{[p}p_2{}^{n]} - 2 z \e_2 p_1{}^{[p}q^{n]} + z \e_1 q^{[p} p_2{}^{n]}\\
    &= 2 (b^{[pn]} + a^{[pn]}) - z \e_2 q^{[p} p_1{}^{n]} - 2 z \e_1 q^{[p}p_2{}^{n]} + 2 z \e_2 q^{[p} p_1{}^{n]} + z \e_1 q^{[p} p_2{}^{n]}\\
    & = 2(b+a)^{[pn]} + z q^{[p} (\e_2 p_1{}^{n]} - \e_1 p_2{}^{n]}) = 2 \b^{pn} + z q^{[p}t^{n]},
\end{aligned}
\end{equation}
where the generator $t = \e_2 p_1 - \e_1 p_2 $ commutes with all other generators. Therefore deformation induced by $t\wedge q$ is simply an additional TsT transformation. The latter has been already show in \cite{Sakamoto:2017cpu} to be a coordinate transformation in the doubled space. 

On principle, one may return to the beginning of the calculation and reorganize $\x$ such as to obtain purely $\b^{pn}$ at the end. We intentionally keep the narrative in this form for two reasons. First, in this form the derivation repeats that of the uni-vector case and poses few questions, such as: what is the reason for it not to work precisely as in the uni-vector case, what is the geometry behind the Yang--Baxter equations appearing here. Second, the necessity of the additional term, corresponding to the TsT transformation generated by $q\wedge t$ is completely unclear, as well as its explicit form. Therefore, more general analysis is needed, probably involving almost-abelian deformations of a different form.

\section{Conclusions}
\label{sec:concl}

In this work we study the geometry behind bi-vector Yang--Baxter deformations. For that we construct an analogue of such deformations, generated by a single Killing vector, which we call uni-vector deformations for homogeneity of notations. These are field transformations in the Einstein--Maxwell dilaton theory, that map solutions into solutions. Explicit transformation rules \eqref{glDdeformedsolution1} look very similar to those of Yang--Baxter deformations, although they cannot be written in the nice form of the open-closed string map. 

The construction of uni-vector deformations follows precisely the logic of poly-vector deformations in supergravity \cite{Bakhmatov:2019dow,Bakhmatov:2020kul,Gubarev:2020ydf}. We considered Einstein--Maxwell dilaton theories that allow a $\rmGL(D+1)$ covariant description in terms of General Relativity on the full $D+1$-dimensional space-time (extended space in the terminology of double field theory). Then the deformations are given by local $\rmGL(D+1)$ transformations parametrized by a Killing vector $\a^m$. We show that the only condition for such a deformation to generate new solutions is that $\a^m$ is Killing. In contrast to bi-vector deformations no analogue of Yang--Baxter equation appears in this formalism. 

In addition to some interesting applications of this solution generating technique to be discussed below, important for us is that uni-vector deformations are equivalent to coordinate transformations in the full $D+1$-dimensional space. In other words, having a complicated form in the lower-dimensional theory these have simple geometrical meaning in the parent theory. We show this by explicit calculations, the coordinate transformation is given by \eqref{eq:unicoord}.

Given much analogy with bi-vector deformations it is tempting to expect that these are also equivalent to coordinate transformations, but in the extended space of double field theory. Indeed, for TsT transformations, at linear level for the general case and explicitly for certain backgrounds this has been shown in \cite{Sakamoto:2017cpu}. Here we manage to show that non-abelian Yang-Baxter bi-vector deformations of special class, called almost abelian in \cite{Borsato:2016ose}, are equivalent to coordinate transformations in the full doubled space. As in the uni-vector case parameter of the transformation explicitly depends on additional coordinates $\tx_m$. New feature is that closure of algebra of such transformations requires classical Yang--Baxter equation if the bi-vector is chosen in the bi-Killing ansatz. This can be though of as the geometric meaning of the classical Yang--Baxter equation in this formalism. 

It is natural to expect that equivalence to coordinate transformations must hold for any bi-vector Yang--Baxter deformations not only almost-abelian. However we found a general proof to be too involved technically and reserved it for future work. If true, this would automatically imply preservation of integrability under bi-vector Yang--Baxter deformation providing both an explicit proof and the underlaying reason why integrability is preserved. Indeed, it is possible to reformulate 2d sigma-model as a doubled sigma model defined on the extended space of double field theory. This was used in \cite{Orlando:2019rjg} to prove that any TsT transformation preserves integrability using its representation as a global $\rmO(10,10)$ transformation of the field of the doubled sigma-model. They show that the Lax connection maps into a Lax connection and hence integrability is preserved in the Liouville sense. If all non-abelian Yang--Baxter transformation are also simply doubled coordinate transformation, this statement would be true in general.

In addition to this intriguing idea plenty of open questions remain. First, we were not able to see the geometric origin of the unimodularity condition. On principle, they could be no any, since breaking of unimodularity maps supergravity backgrounds to those of the generalized supergravity, that can be dualized into a usual supergravity solution by a special coordinate transform inside double field theory (see e.g. \cite{Sakatani:2016fvh}). It is worth to having more detailed explanation of this. Second, one is certainly interested in generalizing the whole picture to the case of tri-vector deformations. For that one would need the analogue of the tensorial transformation law \eqref{eq:largecoordF} for exceptional field theories, and probably an analogue of almost-abelian deformations.

Let us now return to possible applications of uni-vector deformations as solution generating technique in Einstein--Maxwell dilaton theory. To start with, such theory finds useful applications in holographic QCD \cite{Rannu:2021pcq,Arefeva:2024mtl,Lilani:2025wnd}, cosmology \cite{Ghezelbash:2015dka}, particle phenomenology \cite{Raza:2025uda,Junior:2021svb}. In this context generating new solutions appears to be valuable for constructing and investigating moduli space of feasible models. The formalism of uni-vector deformations seems to add to the collection of known solution generating techniques such as \cite{Cadoni:2018pav,Vigano:2022hrg,Yazadjiev:2006ew}. Furthermore, uni-vector deformations might be equivalent to one of the known solution generating methods, that would be an interesting observation. Especially, given their analogy to bi-vector deformations.

\section*{Acknowledgments}

This work has been in part supported by Russian Science Foundation grant RSCF-24-71-10058. We are grateful to Victor Mishnyakov for his valuable comments and  questions.

\appendix
\setcounter{equation}{0}

\section{Coordinate transformations}
\label{app:coord}

The finite translation along a vector field $\x=\x^M\dt_M$ is given by the action of the exponent of the field:
\begin{equation}
X'^M=e^{\x}X^M=\sum_{n=0}^{\infty}\fr{1}{n!}\x^nX^M.
\end{equation}
The recurrent approach gives the desired derivative $\dt_NX'^M$ in the most direct way. Firstly consider a term of the order $n$ in the sum and take its derivative
\begin{equation}
\dt_N(\x^n X^M)=:\F^{(n)}{}_N{}^M.
\end{equation}
The derivative of the next term in the sum reads
\begin{equation}
\begin{split}
&\dt_N(\x^{n+1}X^M)=\dt_N\left(\x^A\dt_A(\x^{n}X^M)\right)=\dt_N\x^A\dt_A(\x^n X^M)+\x^A\dt_A\left(\dt_N(\x^nX^M)\right)=\\
&=a_N{}^A\F^{(n)}{}_A{}^M+\x(\F^{(n)}{}_N{}^M)=\left[(\x \id+a)\F^{(n)}\right]_N{}^M.
\end{split}
\end{equation}
The final step is to calculate first non--trivial $\F^{(n)}$ that is of the order $n=2$
\begin{equation}
\F^{(2)}{}_N{}^M=\dt_N(\x^2 X^M)=\left[(\x \id+a)a\right]_N{}^M=\left[(\x \id+a)^2 {\bf 1}\right]_N{}^M.
\end{equation}
Here ${\bf 1}$ means that the operator act on the constant number and $\x(1)=0$, e.g
\begin{equation}
(\x \id+a)(\x \id +a){\bf 1}=(\x \id+a)a=\x(a)+a^2.
\end{equation} 
In these notations the full sum has the following form
\begin{equation}
\label{Asum}
\dt_NX'^M=\dt_Ne^{\x}X^M=\left(\sum_{n=0}^{\infty}\fr{1}{n!}(\x \id+a)^n{\bf 1}\right)_N^M= \left(e^{(\x \id+a)}{\bf 1}\right)_N{}^M.
\end{equation}

On the other hand we can consider the sum in the \eqref{Asum} as a Taylor of a function of $t$ around $0$ at $t=1$ \cite{Hohm:2012gk}
\begin{equation}
h(t)=\left(\sum_{n=0}^{\infty}\fr{1}{n!}(\x \id+a)^nt^n\right)=e^{t(\x \id+a)}e^{-t\x \id}.
\end{equation}
Indeed, passing from the $n$th derivative to the $n+1$ gives
\begin{equation}
\begin{split}
&h^{(n)}(t)=e^{t(\x \id+a)}(g_n)e^{-t\x \id};\\
&h^{(n+1)}(t)=e^{t(\x \id+a)}(\x \id+a)g_ne^{-t\x \id}.
\end{split}
\end{equation}
This gives $h^{{n}}(0)=(\x+a)^n$ and the Taylor expansion for the function $h(t)$ at $t=1$ has the same form as \eqref{Asum}:
\begin{equation}
h(t)=\sum_{n=0}^{\infty}\fr{1}{n!}h^{(n)}(0)t^n=\left(\sum_{n=0}^{\infty}\fr{1}{n!}(\x \id+a)^nt^n\right).
\end{equation}
Finally we have for the derivative $\dt_NX'^M$ in the matrix notation
\begin{equation}
\label{eq:op2fun}
\fr{\dt X'{}^M}{\dt X^N}=\left(e^{(\x \id+a)}{\bf 1}\right)_N{}^M=\left(e^{(\x \id+a)}\right){}_N{}^Me^{-\x}.
\end{equation}
In the term after the first equality the operators $\x$ act on the constant function $1$ whereas the term after the second equality shows the multiplication of the operators. Due to the construction this multiplication provides an ordinary matrix in the end.

\section{Ricci tensor and Ricci scalar in terms of anholonomicity coefficients}\label{appRicciGTR}

Here we show how to rewrite Ricci tensor and Ricci scalar in terms of anholonomicity coefficients (also referred to as fluxes) defined as usual as $[e_a,e_b] = f_{ab}{}^c e_c$. Since we keep these general and non-constant, $f_{ab}{}^c$ are convenient variables that encode geometric properties of the background. For the vielbein components $e_{a}{}^{m}$, $e^{a}{}_{m}$ we have
\begin{equation}
\begin{aligned}
f_{ab}{}^{c} = -2 e_{a}{}^{m} e_{b}{}^{n} \partial_{[m}e_{n]}{}^{c}, \qquad f_{a} = f_{ab}{}^{b} .
\end{aligned}
\end{equation}
All covariant objects built from vielbeins, such as Riemann and Ricci tensors and Ricci scalar, can be expressed in terms of these fluxes. To show that, we start with the expression for Christoffel symbols
\begin{equation}\label{eq:crist}
\begin{aligned}
\Gamma_{m n}{}^{k} = & \, \frac{1}{2}\, {h}^{k l} ({\partial}_{m}{{h}_{l n}}\,  + {\partial}_{n}{{h}_{l m}}\,  - {\partial}_{l}{{h}_{m n}}\, )  \\
= &  - \frac{1}{2}\, {e}^{a}\,_{m} {e}^{b}\,_{n} {e}^{c}\,_{l} {f}_{a c}\,^{d} {h}_{b d} {h}^{k l} - \frac{1}{2}\, {f}_{a b}\,^{c} {e}_{c}\,^{k} {e}^{a}\,_{m} {e}^{b}\,_{n} - \frac{1}{2}\, {e}^{a}\,_{m} {e}^{b}\,_{n} {e}^{c}\,_{l} {f}_{b c}\,^{d} {h}_{a d} {h}^{k l},
\end{aligned}
\end{equation}
and we obtain the Ricci tensor as
\begin{equation}\label{eq:riccitens}
\begin{aligned}
R_{n l} = & \,  {\partial}_{k}{{\Gamma}_{n l}\,^{k}}\,  - {\partial}_{n}{{\Gamma}_{k l}\,^{k}}\,  + {\Gamma}_{n l}\,^{p} {\Gamma}_{k p}\,^{k} - {\Gamma}_{k l}\,^{p} {\Gamma}_{n p}\,^{k}  \\
= &   - \frac{1}{2}\, {e}^{a}\,_{l} {e}^{b}\,_{n} {f}_{b c}\,^{d} {f}_{a}\,^{c}\,_{d} - \frac{1}{2}\, {\partial}_{m}{{f}_{a}\,^{b}\,_{c}}\,  {e}_{b}\,^{m} {e}^{c}\,_{l} {e}^{a}\,_{n} + \frac{1}{2}\, {e}_{a l} {e}^{b}\,_{n} {f}^{c} {f}_{b c}\,^{a} - \frac{1}{2}\, {e}^{a}\,_{l} {e}^{b}\,_{n} {f}_{a c}\,^{d} {f}_{b d}\,^{c} \\
& - \frac{1}{2}\, {\partial}_{m}{{f}_{a b}\,^{c}}\,  {e}_{c}\,^{m} {e}^{a}\,_{l} {e}^{b}\,_{n} - \frac{1}{2}\, {\partial}_{m}{{f}_{a}\,^{b}\,_{c}}\,  {e}_{b}\,^{m} {e}^{a}\,_{l} {e}^{c}\,_{n} + \frac{1}{2}\, {e}_{a n} {e}^{b}\,_{l} {f}^{c} {f}_{b c}\,^{a} + {\partial}_{n}{{f}_{a}}\,  {e}^{a}\,_{l} \\
& + \frac{1}{2}\, {e}^{a}\,_{l} {e}^{b}\,_{n} {f}_{c} {f}_{a b}\,^{c} + \frac{1}{4}\, {e}_{a l} {e}_{b n} {f}_{c d}\,^{b} {f}^{c d a},
\end{aligned}
\end{equation}
and the Ricci scalar as
\begin{equation}\label{eq:ricciscalar}
\begin{aligned}
R = & \,  h_{n l} ( {\partial}_{k}{{\Gamma}_{n l}\,^{k}}\,  - {\partial}_{n}{{\Gamma}_{k l}\,^{k}}\,  + {\Gamma}_{n l}\,^{p} {\Gamma}_{k p}\,^{k} - {\Gamma}_{k l}\,^{p} {\Gamma}_{n p}\,^{k} )  \\
= & - \frac{1}{2}\, {f}_{a b}\,^{c} {f}_{c d}\,^{a} {h}^{b d} - \frac{1}{4}\, {f}_{a b}\,^{c} {f}_{d f}\,^{g} {h}_{c g} {h}^{a d} {h}^{b f} + {\partial}_{m}{{f}_{a}}\,  {e}_{b}\,^{m} {h}^{a b} - {f}_{a} {f}_{b} {h}^{a b} + {\partial}_{m}{{f}_{a}}\,  {e}^{a}\,_{n} {h}^{m n}.
\end{aligned}
\end{equation}

\bibliography{bib.bib}
\bibliographystyle{utphys.bst}

\end{document}

%% file: Defs.tex
\def\a{\alpha} \def\b{\beta} \def\g{\gamma} \def\d{\delta} \def\e{\epsilon}
  \def\h{\eta} \def\q{\theta}
  \def\k{\kappa} \def\l{\lambda} 
 \def\x{\xi}   \def\r{\rho}
 \def\s{\sigma}   \def\f{\phi}
\def\vf{\varphi}

\def\D{\Delta} 
   
   \def\L{\Lambda}

\def\F{\Phi}   \def\W{\Omega}

\def\fr{\frac}  \def\dt{\partial}

\def\ph{\phantom}
\def\mc{\mathcal}
\def\mA{\mathcal{A}}
\def\mB{\mathcal{B}}
\def\mC{\mathcal{C}}
\def\mD{\mathcal{D}}
\def\mE{\mathcal{E}}
\def\mF{\mathcal{F}}

\def\mH{\mathcal{H}}
\def\mK{\mathcal{K}}
\def\mL{\mathcal{L}}
\def\mM{\mathcal{M}}
\def\mN{\mathcal{N}}
\def\mP{\mathcal{P}}
\def\mQ{\mathcal{Q}}

\def\tx{\tilde{x}}
\def\ty{\tilde{y}}
\def\tz{\tilde{z}}

\def\tdt{\tilde{\partial}}

\def\XX{\mathbb{X}}

\def\SS{\mathbb{S}}

\def\rmGL{\mathrm{GL}}

\def\rmO{\mathrm{O}}
\def\rmSO{\mathrm{SO}}

\def\id{\mathbb{1}}

\newcommand\bqa {\begin{eqnarray}}
\newcommand\eqa {\end{eqnarray}}

\newcommand{\bear}{\begin{array}}
\newcommand{\enar}{\end{array}}

\newcommand{\be}{\begin{equation}}
\newcommand{\ee}{\end{equation}}
\newcommand{\bea}{\begin{eqnarray}}
\newcommand{\eea}{\end{eqnarray}}

\def\rmSO{\mathrm{SO}}
\def\rmGL{\mathrm{GL}}

\def\AdS{\mathrm{AdS}}

\def\td{\tilde{\d}}